\documentclass{aa}

\usepackage{graphicx, psfig, times, amsfonts}

\newcommand{\less}{\raisebox{-1.1mm}{$\stackrel{<}{\sim}$}}
\newcommand{\more}{\raisebox{-1.1mm}{$\stackrel{>}{\sim}$}}
\newcommand{\msol}{\mbox{M$_{\odot}$}}

\newcommand{\msolyr}{{M$_{\odot}$}\,yr$^{-1}$}
\newcommand{\lsol}{\mbox{L$_{\odot}$}}

\newcommand{\M}{{\sc 2mass}}
\newcommand{\OG}{{\sc ogle}}
\newcommand{\DE}{{\sc denis}}

\begin{document}

\title{Long Period Variables in the Magellanic Clouds: \\ \OG\ + \M\ + \DE\
\thanks{
Tables~2, 3, 6, 7 and 8 are available in electronic form
at the CDS via anonymous ftp to cdsarc.u-strasbg.fr (130.79.128.5)
or via http://cdsweb.u-strasbg.fr/cgi-bin/qcat?J/A+A/.
Figures~2, 9 and 13 are available in the on-line edition of A\&A.}
}

\author{
M.A.T. Groenewegen
}

\institute{
Instituut voor Sterrenkunde, K.U. Leuven, PACS-ICC, Celestijnenlaan 200B, 
B--3001 Leuven, Belgium 
}

\date{received: 2004,  accepted: 2004}

\offprints{Martin Groenewegen (groen@ster.kuleuven.ac.be)}


\abstract{
The 68~000 $I$-band light curves of variable stars detected by the
\OG\ survey in the Large and Small Magellanic Clouds (MCs) are fitted
by Fourier series, and also correlated with the \DE\ and \M\ {\it
all-sky release} databases and with lists of spectroscopically
confirmed M-, S- and C-stars. Lightcurves and the results of the
lightcurve fitting (periods and amplitudes) and \DE\ and \M\ magnitudes are
presented for 2277 M-,S-,C-stars in the MCs. The following aspects are
discussed: the $K$-band period-luminosity relations for the
spectroscopically confirmed AGB stars, period changes over a timespan
of about 17 years in a subset of about 400 LPVs, and candidate
obscured AGB stars.
The use of a sample of spectroscopically confirmed variables allows me
to show specifically that almost all carbon stars are brighter than
the tip of the RGB, and occupy sequences A+,B+,C and D.
It is shown (for the LMC where there is a sufficient number of
spectroscopically identified M-stars) that for sequences A+,B+,C the
M-stars are on average fainter than the C-stars, as expected from an
evolutionary point of view and previously observed in MC
clusters. However, this is not so for sequence ``D'', suggesting that
the origin of the so-called Long Secondary Periods is not related to
an evolutionary effect. The fraction of objects that has a period
on sequence ``D'' is also independent of chemical type.
Three stars are identified that have been classified as oxygen-rich in
the 1970s and carbon-rich in 1990s. Possibly they underwent a thermal
pulse in the last 20 years, and dredged-up enough carbon to switch
spectral type.
The observations over almost two decades seem to suggest that up to
10\% of AGB variables changed pulsation mode over that time span. More
robust estimates will come from the ongoing and future (microlensing)
photometric surveys.
A sample of 570 variable red objects ($(J-K) >$ 2.0 or $(I-K) >$ 4.0)
is presented in which most stars are expected to be dust obscured AGB stars.
Estimates are presented for cut-offs in $(J-K)$ which should be applied to
minimise dust obscuration in $K$, and based on this, C- and O-star
$K$-band $PL$-relations for large amplitude variables in the SMC and
LMC are presented.
\keywords{Stars: AGB and post-AGB - Stars: carbon - Stars: variables:
general - Magellanic Clouds} }


\maketitle

\section{Introduction}

In the course of the micro lensing surveys in the 1990's, the
monitoring of the Small and Large Magellanic Clouds has revealed an
amazing number and variety of variable stars. A big impact was felt
and is being felt in many areas of variable star research, like Cepheids
and RR Lyrae stars. Also in the area of Long Period Variables
(LPVs) and AGB stars there has been remarkable progress. Wood et
al. (1999) and Wood (2000) were the first to identify and label
different sequences ``ABC'' thought to represent the classical Mira
sequence (``C'') and overtone pulsators (``A,B''), and sequence ``D''
which is still unexplained (Olivier \& Wood 2003, Wood 2003, 2004). Stars
on these sequence are sometimes referred to as having Long Secondary
Periods--LSPs. This view has subsequently been confirmed and expanded
upon by Noda et al. (2002), Lebzelter et al. (2002), Cioni et
al. (2003), Ita et al. (2004a,b) and Kiss \& Bedding (2003, 2004). 
These works differ in the source of the variability data ({\sc macho},
\OG, {\sc eros}, {\sc moa}), area (SMC or LMC), associated infrared
data (Siding Spring 2.3m, \DE, \M, {\sc sirius}), and selection on
pulsation amplitude or infrared colours.

The present paper considers the \OG\ data for both LMC and SMC. Ita et
al. (2004a) only consider the \OG\ data 
in overlap with their {\sc sirius} IR observations in LMC and SMC, and
Kiss \& Bedding consider only stars in the SMC with \M\ data with $(J-K) > 0.9$.

Also in contrast to previous studies, emphasis is put on
spectroscopically confirmed AGB stars (i.e. M-, S- and C-stars). In
other studies M- and C-stars are usually identified photometrically by
using a division at a $(J-K)$ colour of $\sim$1.4 mag. This paper
specifically addresses the properties of known carbon stars in
relation to sequences ``ABCD''.

The paper is structured as follows. In Section~2 the \OG, \M\ and \DE\
surveys are described. In Section~3 the model is presented, both in
terms of the actual lightcurve fitting, and the post-processing.
Section~4 presents the results. Discussed are the $K$-band
$PL$-relation for the spectroscopically confirmed AGB stars, period
changes over a timespan of almost 2 decades, and a sample of very red
obscured AGB star candidates. The conclusions are summarised in
Section~5.  Some of this work, and the star-to-star comparison of
periods derived by me from {\sc macho} and \OG\ data and literature
values are described in Groenewegen (2004).

\section{The data sets}

Zebru\'n et al. (2001) describe the dataset of the about 68~000
variable objects detected by \OG\ in the direction of the LMC and SMC,
obtained in the course of the {\sc ogle-ii} micro lensing survey using the
difference image analysis (DIA) technique. Twenty-one fields in the
central parts of the LMC, and 11 fields in the central parts of the
SMC of size 14.2\arcmin$\times$57\arcmin\ each were observed in $BVI$,
with an absolute photometric accuracy of 0.01-0.02 mag.  The large
majority of data was obtained in the $I$-band (and the DIA analysis
has been done only on the $I$-band data), and these data were
downloaded from the \OG\ homepage (http://sirius.astrouw.edu.pl/$^{\sim}$ogle/).

The \DE\ survey is a survey of the southern hemisphere in $IJK_{\rm
s}$ (Epchtein et al. 1999). Cioni et al. (2000) describe a point
source catalog of sources in the direction of the Magellanic Clouds
(MCs; DCMC = \DE\ Catalogue towards the Magellanic Clouds) containing
1~635~680 objects with a detection in at least 2 of the 3 photometric
bands that is available in electronic form. The 68193 \OG\ objects
were correlated on position using a 3\arcsec\ search radius and 40793
matches were found.

The \M\ survey is an  all-sky survey in the $JHK_{\rm s}$ near-infrared
bands. On March 25, 2003 the \M\ team released the all-sky point
source catalog. The easiest way to check if a star is included in the
\M\ database is by uplinking a source table with coordinates to the
\M\ homepage. Such a table was prepared for the 68193 \OG\ objects and
correlated on position using a 3\arcsec\ radius. Data on 50129 objects
were returned.

\begin{table*}
\setlength{\tabcolsep}{1.3mm}

\caption{Comparison of coordinates, and number of positional matches, before and after a correction was applied.}
\begin{tabular}{lrrrrrrl} \hline
\small
OGLE- ({\it other}) & $\Delta$ RA $\cos (\delta)$ & $\Delta \delta$ & $N$ &  $\Delta$ RA $\cos (\delta)$ & $\Delta \delta$ & $N$ & remark \\ \hline 
MACHO               &   0.52 $\pm$ 0.61  &   0.21 $\pm$  0.63 &  647 &  0.06 $\pm$ 0.63  &   0.05 $\pm$  0.64 &  662 & mainly ``blue objects'' and EBs \\
MACHO               &   0.52 $\pm$ 1.28  &   0.14 $\pm$  1.23 & 1427 &  0.11 $\pm$ 1.31  & $-0.02$$\pm$  1.22 & 1436 & LPVs; Wood et al. 1999 \\

MOA                 & $-0.01$$\pm$ 0.32  &   0.008 $\pm$ 0.55 &  123 & $-$  &    $-$ &  $-$ & EBs; no correction applied \\
MOA                 &   0.78 $\pm$ 1.04  &   0.002 $\pm$ 0.94 &  150 &  0.11 $\pm$ 0.97  &   0.03 $\pm$  0.94 &  150 &  LPVs; Noda et al. 2002 \\

EROS                &   0.41 $\pm$ 0.50  & $-0.21$$\pm$  0.48 &  368 &  0.06 $\pm$ 0.58  &   0.01 $\pm$  0.49 &  381 & cepheids and EBs \\
AGAPEROS            &   0.31 $\pm$ 1.09  &   0.37 $\pm$  1.00 &  503 &  0.05 $\pm$ 1.07  &   0.05 $\pm$  0.95 &  501 & LPVs; Letzelter et al. 2002 \\

DENIS               & $-0.10$$\pm$ 1.33  &   0.02 $\pm$  1.37 &  313 & $-$  &    $-$ &  $-$ & LPVs; Cioni et al. 2003; no correction applied \\

HUGHES              &   2.19 $\pm$ 1.50  &   0.85 $\pm$  1.41 &  433 &  0.22 $\pm$ 1.34  & $-0.01$$\pm$  1.34 &  436 & LPVs \\
KDM                 &   0.55 $\pm$ 1.12  & $-0.23$$\pm$  0.90 & 1024 &  0.06 $\pm$ 1.11  & $-0.03$$\pm$  0.90 & 1028 & carbon stars \\
RAW                 & $-1.06$$\pm$ 0.80  &   1.03 $\pm$  0.71 &  873 &  0.04 $\pm$ 0.76  &   0.04 $\pm$  0.70 &  874 & carbon stars \\
BMB                 &   0.42 $\pm$ 0.90  &   0.00 $\pm$  0.66 &  381 &  0.05 $\pm$ 0.96  & $-0.01$$\pm$  0.68 &  385 & AGB stars \\

\hline
\end{tabular}
\label{tab-astro}
\end{table*}

\section{The model}

At the heart of the data processing are two numerical codes, that are
described in detail in the Appendices. Briefly, the first code (see
for details Appendix A) sequentially reads in the $I$-band data for
the 68~000 objects, determines periods through Fourier analysis, fits
sine and cosine functions to the light curve through linear
least-squares fitting and makes the final correlation with the
pre-prepared \DE\ and \M\ source lists. All the relevant output
quantities are written to file.

This file is read in by the second code (see for details Appendix B). 
A further selection may be applied (typically on period, amplitude and
mean $I$-magnitude), multiple entries are filtered out (i.e. objects
that appear in different \OG\ fields), and a correlation is made with
pre-prepared lists of known non-LPVs and known LPVs or AGB stars. The
output of the second code is a list with LPV candidates.

The third step (for details see Appendix C) consists of a visual
inspection of the fits to the light curves of the candidate LPVs and a
literature study through a correlation with the {\sc simbad}
database. Non-LPVs are removed, and sometimes the fitting is redone. 
The final list of LPV candidates is compiled.

\begin{figure}
\vskip6cm
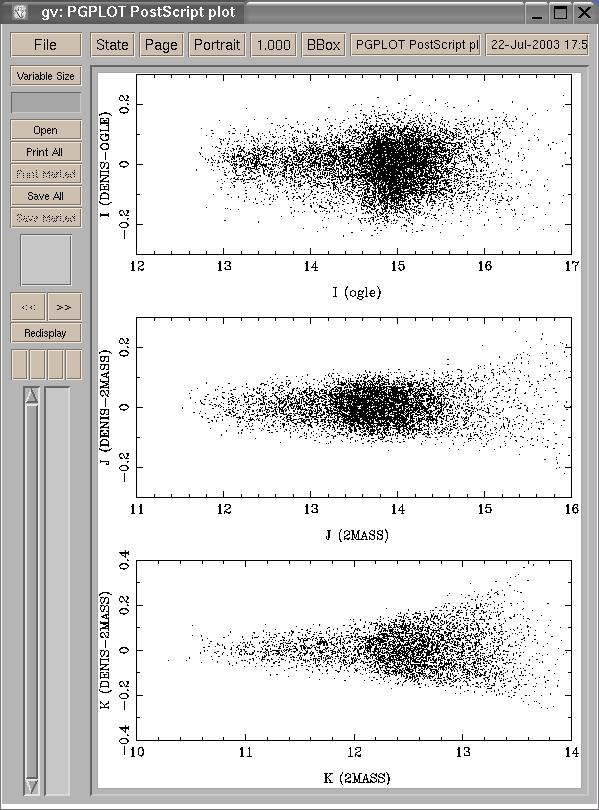
\vskip5.5cm
\caption[]{
Difference in photometry, {\em after} the following offsets have been
applied: $I$(denis-ogle) = $-0.018$, $J$(denis-2mass)= $-0.090$,
$K$(denis-2mass)= $-0.14$. Used in the analysis are the stars with an
$I$-band amplitude $<0.05$ mag. Plotted are about 12000, 8700, and
5700 stars, top to bottom.
}
\label{Fig-PHT}
\end{figure}

\begin{table*}
\caption[]{
First entries in the electronically available table, which lists:
OGLE-field, OGLE-name, the three fitted periods with errors and
amplitude (0.00 means no fit), mean $I_{\rm ogle}$, and associated \DE\
$IJK$ photometry with errors, and associated \M\ $JHK$ photometry with
errors (99.9 and 9.99 means no association, or no value).
}
\includegraphics[angle=+90, width=190mm]{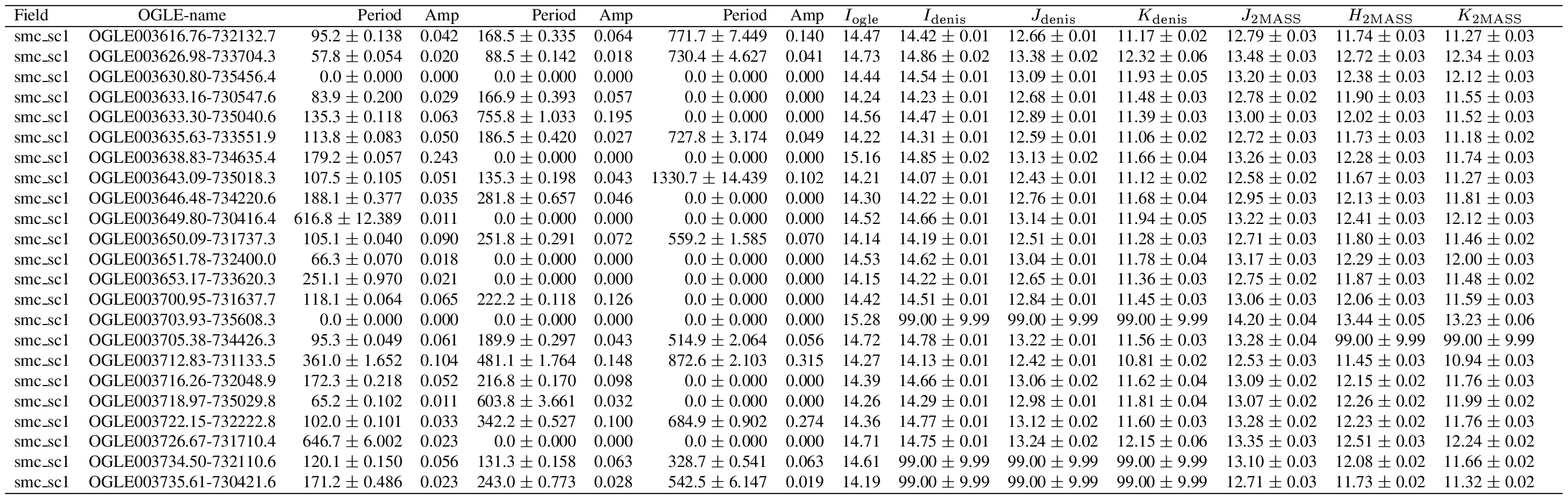}

\label{TAB-A}
\end{table*}

\begin{table}
\caption[]{
First entries in the electronically available table, which list:
OGLE-field, OGLE-name, other names, spectral type and references and
comments.
}
\includegraphics[angle=+90]{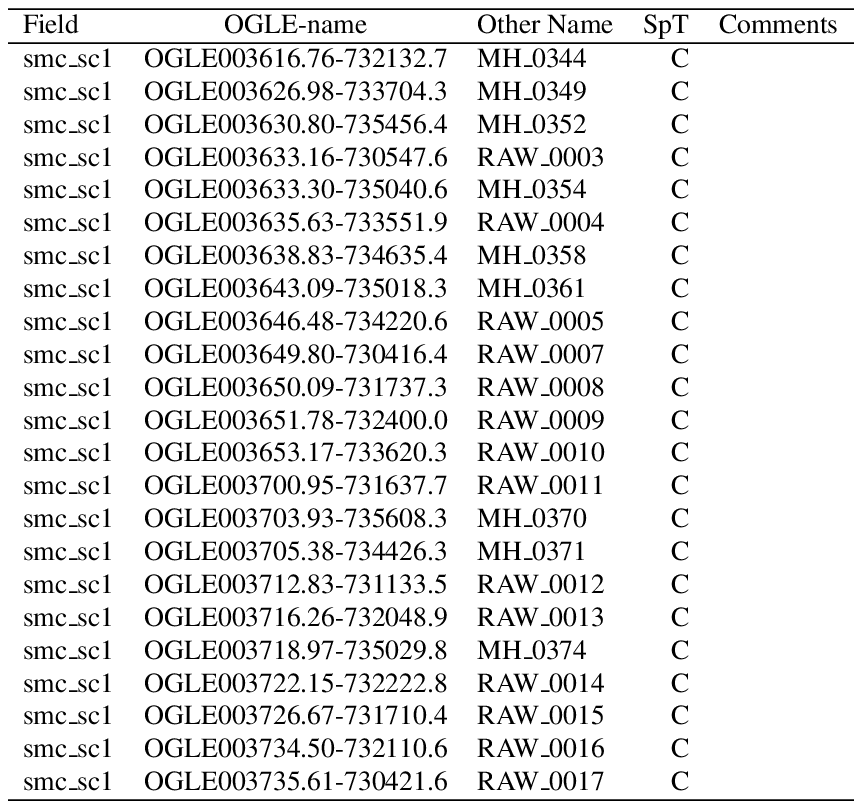}

\label{TAB-B}
\end{table}

\section{Results}

\subsection{Astrometry}

The spatial correlation between the OGLE objects and known LPVs and
AGB stars, and known non-LPVs, is actually done in 2 steps. In the
first step the correlation is made, and the differences and spread in
$\Delta$RA $\cos (\delta)$ and $\Delta \delta$ are determined. These
mean offsets are then applied to make the final cross-correlation.
The results are listed in Table~\ref{tab-astro}. Regarding the MACHO
data the agreement between the offsets determined from the list of
non-LPVs, and the data by Wood et al. (1999) is good and a combined
offset of $\Delta$ RA $\cos (\delta) = 0.52 \arcsec$ and $\Delta
\delta = 0.16 \arcsec$ is applied.  With respect to the other sources
of astrometry, similar small changes have been found, and in most
cases applied, usually increasing the number of matches in the final
matching.

\subsection{Photometry}

As was already discussed in Groenewegen (2004), by selecting the least
variable stars in the \OG\ database one can compare photometry.

In particular the \OG\ $I$ was compared to the (singe-epoch) \DE\ $I$,
and the (single-epoch) \DE\ $JK$ was compared to the (single-epoch)
\M\ $JK$ magnitudes. This was done by selecting those objects with an
amplitude in the $I$-band of $<0.05$ mag. 

Figure~\ref{Fig-PHT} shows the final results when offsets
$I$(denis-ogle) = $-0.018$, $J$(denis-2mass)= $-0.090$, and
$K$(denis-2mass)= $-0.14$ are applied.  The offsets derived here are
very similar to those derived in Delmotte et al. (2002) using a
similar analysis based on a direct comparison of \DE\ with the \M\ 2nd
incremental data-release (they found: $J$(denis-2mass) = $-0.11 \pm
0.06$, and $K$(denis-2mass)= $-0.14 \pm 0.06$).

\subsection{The sample of spectroscopically confirmed M,S,C-stars}

\begin{figure*}
\vskip8cm
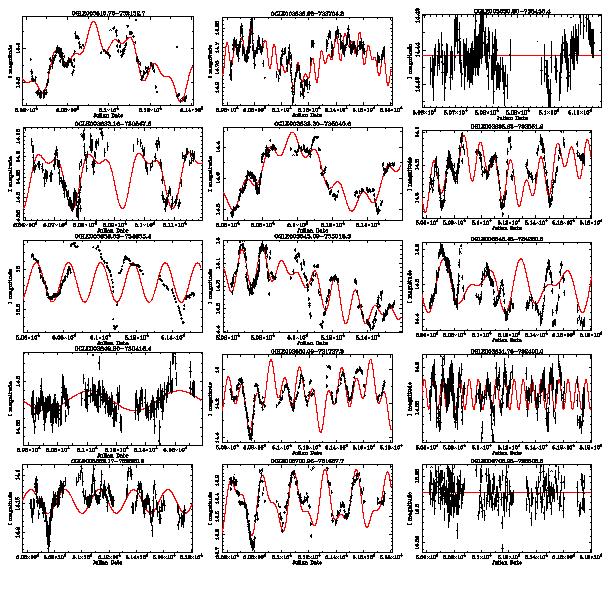
\vskip7cm
\caption[]{
First entries of electronically available figure with all
lightcurves. The fit is indicated by the (red) solid line. Crosses
indicate data points not included in the fit.
}
\label{Fig-LC}
\end{figure*}

As mentioned before, a correlation was made with a list that contains
12631 objects over the entire MCs with a spectroscopically determined
spectral type. Using a 4\arcsec\ search radius, 2478 unique objects
were found in the \OG\ fields.  After visual inspection (removing
mostly obviously incorrect positional mismatches, indicated by
association with very faint \OG\ objects with hardly any variability
[sect.~3.2.2]) there remain 2277 objects (856 C- and 3 M-stars in the
SMC, 1064 C-, 10 S-, 344 M-stars in the LMC) which are the subject of
further study. Their lightcurves are shown in Figure~\ref{Fig-LC}
(only the first few, the full figure only being available in the
electronic edition), the results of the lightcurve analysis and
association with \DE\ and \M\ is given in Table~\ref{TAB-A}, and the
association with known objects and additional comments and references
are given in Table~\ref{TAB-B} (again, only the first few entries are
shown, the full tables being available only electronically).

In the discussion that follows, magnitudes are dereddened using the
$A_{\rm V}$ values that correspond to the respective OGLE field in the
SMC or LMC, and selective reddenings of $A_{\rm I}/A_{\rm V} = 0.49$,
$A_{\rm J}/A_{\rm V} = 0.27$, $A_{\rm H}/A_{\rm V} = 0.20$, 
$A_{\rm K}/A_{\rm V} = 0.12$ (Draine 2003) are used.

Some of the objects have different spectral classifications in
different surveys. These have been identified in Table~\ref{TAB-B}.
Some of them are in fact due to a typographical error in CML (Cioni,
private communication). In all other cases the different sources of
photometry, pulsation periods (when available) and close proximity
suggest that these objects are one and the same. For one object,
OGLE052711.00-692827.5, both associated objects have an uncertain
spectral type: WBP-48 is classified as (M?), SHV052733.4-693050 as
(CS?).  It is kept as ``M''.  OGLE052705.07-693606.9 is associated
with the carbon star KDM-4094, but also with the M-star HV 12048, and
the latter type is assumed here. Of course oxygen-rich stars may
evolve into carbon-rich stars, and three more objects that may have
evolved in this way judging from the different spectral types are
OGLE050859.83-691458.0 (WOH-G-202, KDM-2226), OGLE054109.00-700942.1
(WOH-G-473, KDM-5626), and OGLE054120.38-700823.3 (WOH-G-478,
KDM-5645).  In all cases the more recent spectral type (C) has been
adopted. These stars are valid targets for spectroscopy, both in the
optical to determine their ($s$-process) abundances, and in the
mid-infrared to identify the dust features of the possibly still
oxygen-rich dust shell.

Six objects are erroneously associated by SIMBAD with relatively
bright objects, and an appropriate comment is added in
Table~\ref{TAB-B}. The alledged counterparts have $B$ and/or $V$
magnitudes in the range 10-13 mag, while the $I_{\rm ogle}$, and the
photo-electric $R$ and $I$ magnitude (all six objects happen to be
listed in KDM) are typically 14-15 mag and suggest $B,V$ magnitudes
that would be closer to 16-18 mag.  Optical and \M\ finding charts
were also inspected and a bright object was typically found within
1\arcmin\ of the \OG\ object.

One of the major finding by Wood et al. (1999), Wood (2000) and
subsequently confirmed by other studies is the existence of distinct
$PL$-relations, usually indicated by the letters A-,A+,B-,B+,C,D. The
values for the boundaries of these regions were originally taken from
Ita et al. (2004a), Ita (private communication) and then
slightly adapted. They are basically drawn by eye, and for reference
the lines that define the boundaries have been listed in
Table~\ref{tab-seq}. The boundaries between A- and A+, and B- and B+,
are placed at the tip of the RGB, i.e. $K_0$ = 12.1, and 12.7 in the
LMC and SMC, respectively.

\begin{table*}
\caption{Definition of the boundary lines of the sequences in the
$K$-band $PL$-diagram in Figure~\ref{fig:PL}. Relations are of the form
$K_{\rm min,max}= {\rm slope_{min,max}} \log P + {\rm zp_{min,max}}$,
with $K$ the dereddened magnitude on the \M\ system.}
\begin{tabular}{lrrrrrrl} \hline
 & $K_{\rm max}$ & $K_{\rm min}$ & slope$_{\rm min}$ & 
              zp$_{\rm min}$ & slope$_{\rm max}$ & zp$_{\rm max}$ & comment \\
\hline
\multicolumn{8}{c}{SMC} \\
A- & 12.7 & 13.5 & -3.35 & 17.23 & -3.35 & 17.85 & slope from Ita (priv. comm.) \\
A+ & 10.0 & 12.7 & -3.45 & 17.10 & -3.45 & 17.85 & slope from Ita (priv. comm.) \\
B- & 12.7 & 13.5 & -3.35 & 17.85 & -3.35 & 18.77 & slope from Ita (priv. comm.) \\
B+ & 10.0 & 12.7 & -3.45 & 17.95 & -3.45 & 19.00 & slope from Ita (priv. comm.) \\
C  & 10.0 & 12.7 & -3.85 & 20.05 & -3.85 & 21.45 & slope from Ita (priv. comm.) \\
D  & 10.5 & 13.0 & -3.85 & 21.55 & -3.85 & 23.50 & slope from Ita (priv. comm.) \\
\multicolumn{8}{c}{LMC} \\
A- & 12.1 & 13.5 & -3.35 & 16.70 & -3.35 & 17.50 & slope from Ita (priv. comm.) \\
A+ &  9.4 & 12.1 & -3.45 & 16.50 & -4.00 & 18.45 &  \\
B- & 12.1 & 13.5 & -3.35 & 17.50 & -3.35 & 18.50 & slope from Ita (priv. comm.) \\
B+ &  9.4 & 12.1 & -4.00 & 18.55 & -4.00 & 19.80 & \\
C  &  9.4 & 12.5 & -3.80 & 19.50 & -3.80 & 20.90 & \\
D  &  9.8 & 13.0 & -3.85 & 21.20 & -3.85 & 23.90 & slope from Ita (priv. comm.) \\
\hline
\end{tabular}
\label{tab-seq}
\end{table*}

Figure~\ref{fig:PL} shows the $PL$-relation in the $K$-band with these
boundaries, for the SMC and LMC separately, and according to cuts in
the $I$-band amplitude, as indicated in the insets. The $K$-band is on
the \M\ system, and is the average of the \DE\ and \M\ photometry. In
particular, if both \DE\ and \M\ $K$-band data is available, the \DE\
data point is corrected as explained above (i.e. 0.14 mag added), and
averaged with the \M\ data point. This should take out some of the
scatter in the $PL$-diagram, as the effect of the variability in the
$K$-band is reduced. If only \DE\ is available, the corrected value is
used.  Not all periods listed in Table~\ref{TAB-A} are plotted.
Inspecting the outliers in preliminary versions of Figure~\ref{fig:PL}
indicated that to reduce the scatter a further culling was necessary,
in particular at the longer periods, as the timespan of the \OG\
observations is about 1100 days. The following conditions were applied
(with $\Delta P$ the error in the period): $\Delta P/P < 0.01$ for $P<
500^d$; $\Delta P < 5^d$ for $500^d < P < 800^d$ and $\Delta P <
1.5^d$ for $P > 800^d$. Figures~\ref{fig:PL}, \ref{Fig-HistoSeq-LMC},
\ref{Fig-HistoSeq-SMC}, \ref{Fig-Amp}, \ref{Fig-PerCol},
\ref{fig-pratchange}, \ref{Fig-CC}, \ref{fig:PLIR} and all
calculations are based on periods that obey these conditions. In
contrast, all periods found have been listed in Tables~\ref{TAB-A} and
\ref{TAB-E} and are shown in Figures~\ref{Fig-LC}, \ref{Fig-LC-LPV}
and \ref{Fig-LC-IR}.

The following systematics may be observed: (1) small amplitude
variables are present at all periods, (2) objects with $I$-band
amplitudes \more 0.05 are not found in box ``A+'' (nor ``A-''), (3)
objects with $I$-band amplitudes \more 0.45 are mainly found in box ``C''.
These three remarks are valid for SMC and LMC alike.

\begin{figure*}[ht]
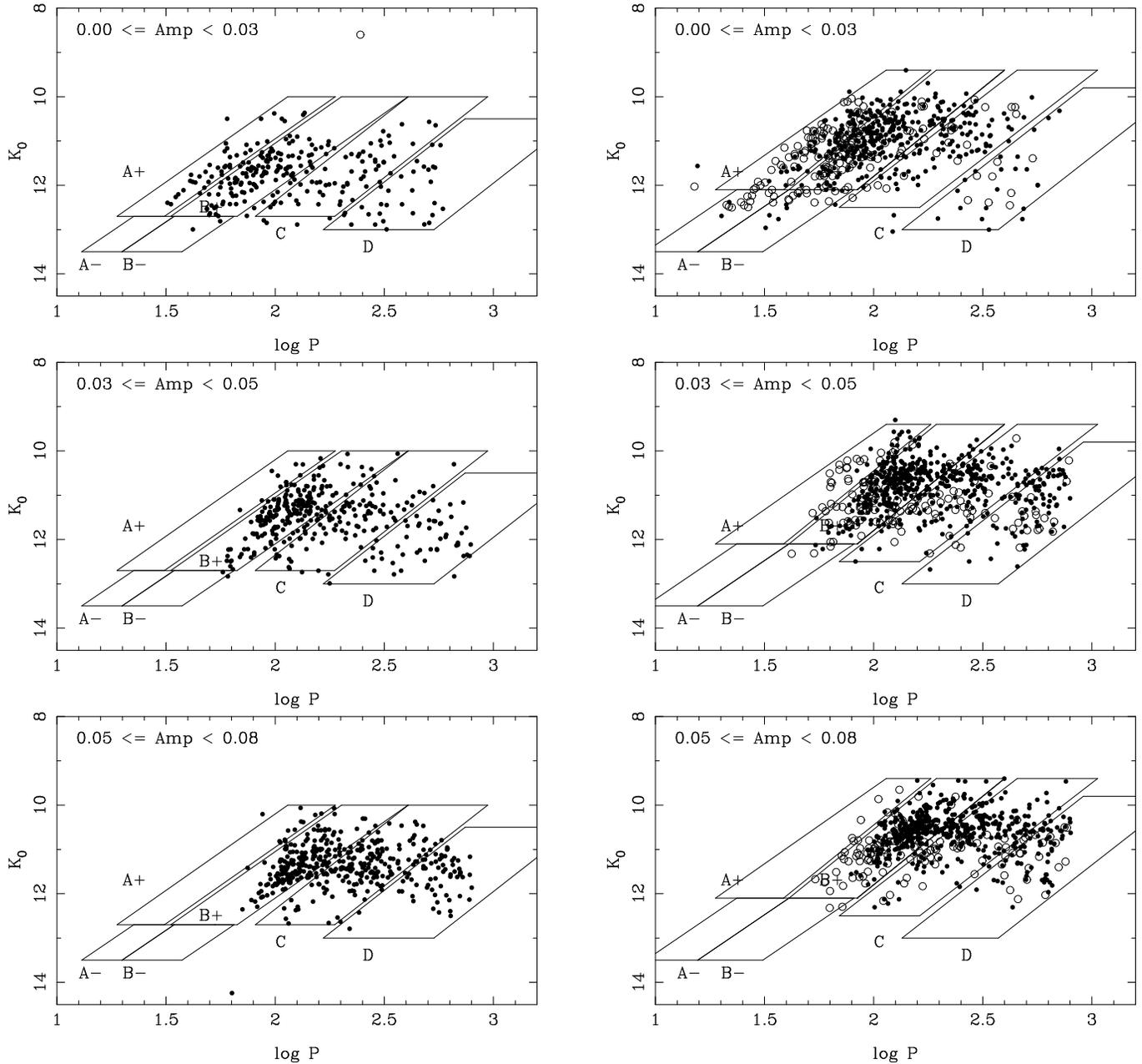

\begin{minipage}{0.47\textwidth}
\resizebox{\hsize}{!}{\includegraphics{K-P_SMC_0.00_0.03.ps}}
\end{minipage}
\hfill
\begin{minipage}{0.47\textwidth}
\resizebox{\hsize}{!}{\includegraphics{K-P_LMC_0.00_0.03.ps}}
\end{minipage}

\begin{minipage}{0.47\textwidth}
\resizebox{\hsize}{!}{\includegraphics{K-P_SMC_0.03_0.05.ps}}
\end{minipage}
\hfill
\begin{minipage}{0.47\textwidth}
\resizebox{\hsize}{!}{\includegraphics{K-P_LMC_0.03_0.05.ps}}
\end{minipage}

\begin{minipage}{0.47\textwidth}
\resizebox{\hsize}{!}{\includegraphics{K-P_SMC_0.05_0.08.ps}}
\end{minipage}
\hfill
\begin{minipage}{0.47\textwidth}
\resizebox{\hsize}{!}{\includegraphics{K-P_LMC_0.05_0.08.ps}}
\end{minipage}
\caption{
$K$-band $PL$-relation, for the SMC (left) and LMC (right). Panles
indicate selection on $I$-band amplitude. Carbon stars are indicated
by filled circles, M- and S-stars by open circels. All periods from
Table~\ref{TAB-A} that fulfil $\Delta P/P < 0.01$ for $P< 500^d$;
$\Delta P < 5^d$ for $500^d < P < 800^d$ and $\Delta P < 1.5^d$ for $P
> 800^d$ are plotted. Boxes related to the ``ABCD'' sequences are
indicated. The functional dependence is summarised in Table~\ref{tab-seq}.
}
\label{fig:PL}
\end{figure*}

\setcounter{figure}{2}
\begin{figure*}[ht]
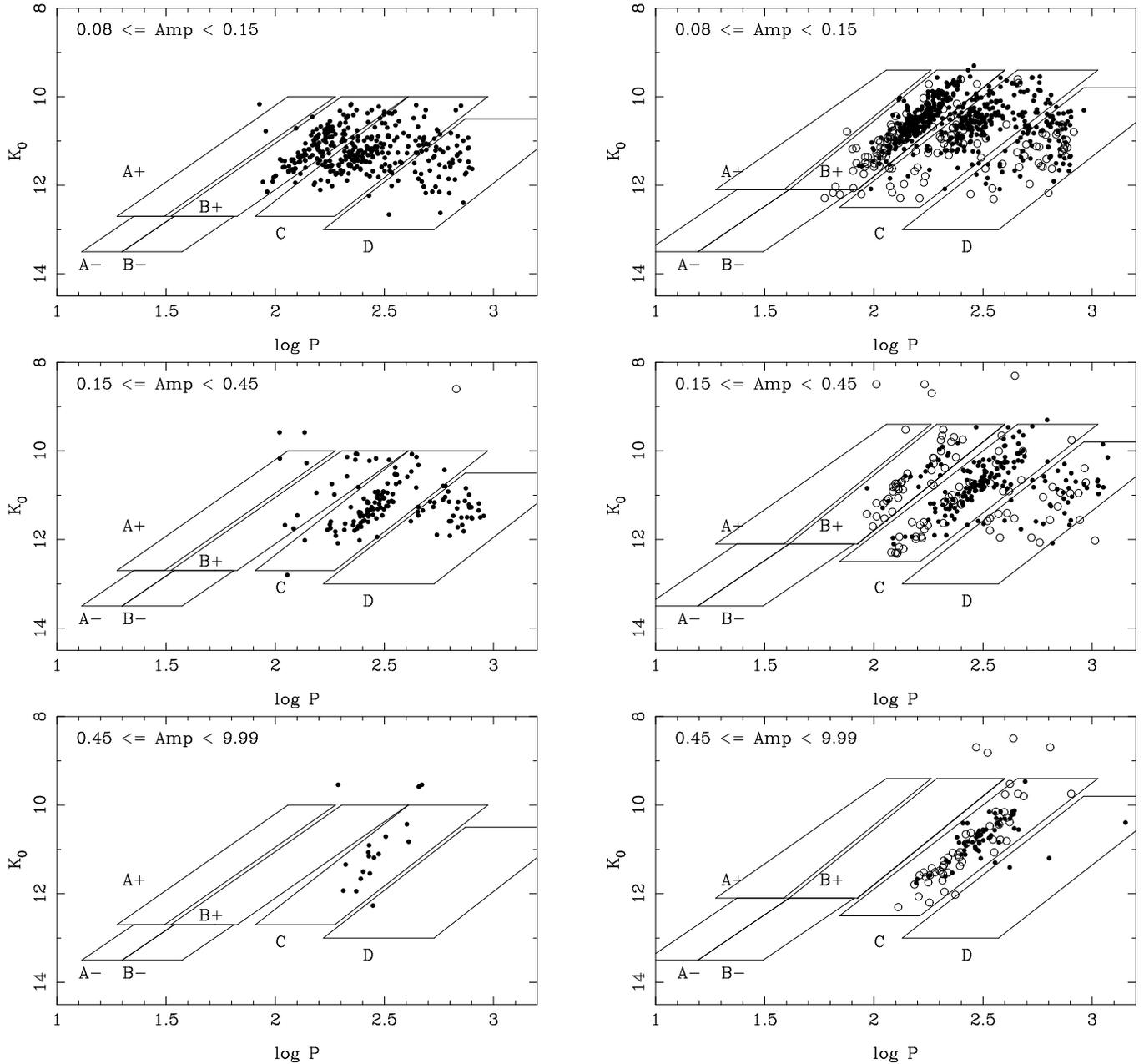

\begin{minipage}{0.47\textwidth}
\resizebox{\hsize}{!}{\includegraphics{K-P_SMC_0.08_0.15.ps}}
\end{minipage}
\hfill
\begin{minipage}{0.47\textwidth}
\resizebox{\hsize}{!}{\includegraphics{K-P_LMC_0.08_0.15.ps}}
\end{minipage}

\begin{minipage}{0.47\textwidth}
\resizebox{\hsize}{!}{\includegraphics{K-P_SMC_0.15_0.45.ps}}
\end{minipage}
\hfill
\begin{minipage}{0.47\textwidth}
\resizebox{\hsize}{!}{\includegraphics{K-P_LMC_0.15_0.45.ps}}
\end{minipage}

\begin{minipage}{0.47\textwidth}
\resizebox{\hsize}{!}{\includegraphics{K-P_SMC_0.45_9.99.ps}}
\end{minipage}
\hfill
\begin{minipage}{0.47\textwidth}
\resizebox{\hsize}{!}{\includegraphics{K-P_LMC_0.45_9.99.ps}}
\end{minipage}
\caption{Continued}
\end{figure*}

Unfortunately, only three confirmed oxygen-rich SMC stars appear in
Tables~\ref{TAB-A} and \ref{TAB-B}, and all come from an IRAS selected
sample (GB98). They all have (very) long periods and the error in
these periods are such that only one period appears in Figure~\ref{fig:PL}. 

For the LMC the situation is better.  One can observe that: (4)
essentially all SMC, and almost all LMC, carbon stars are brighter
than the tip of the RGB, (5) LMC M-stars are observed below the TRGB,
but only at small amplitudes, (6) for a given ``box'' or amplitude
cut, the M-stars are on average fainter than the C-stars, (7) a few
stars brighter than the expected tip of the AGB, presumably
supergiants, are present, and they predominantly have large pulsation amplitudes.

Statement 6 is illustrated in more detail in Figures~\ref{Fig-HistoSeq-LMC} 
and \ref{Fig-HistoSeq-SMC}, where, respectively, LMC and SMC $K$-band
histograms are shown for the stars inside different ``boxes'' and
partly for different $I$-band amplitude cuts, as indicated in the
insets. For the LMC a distinction is made between C- and M/S-stars and
for sequences A+,B+ and C the M-stars are on average fainter than the C-stars. 
The seems not to be the case for sequence D, suggesting that
the LSP phenomenon is not related to evolutionary status.

\begin{figure*}
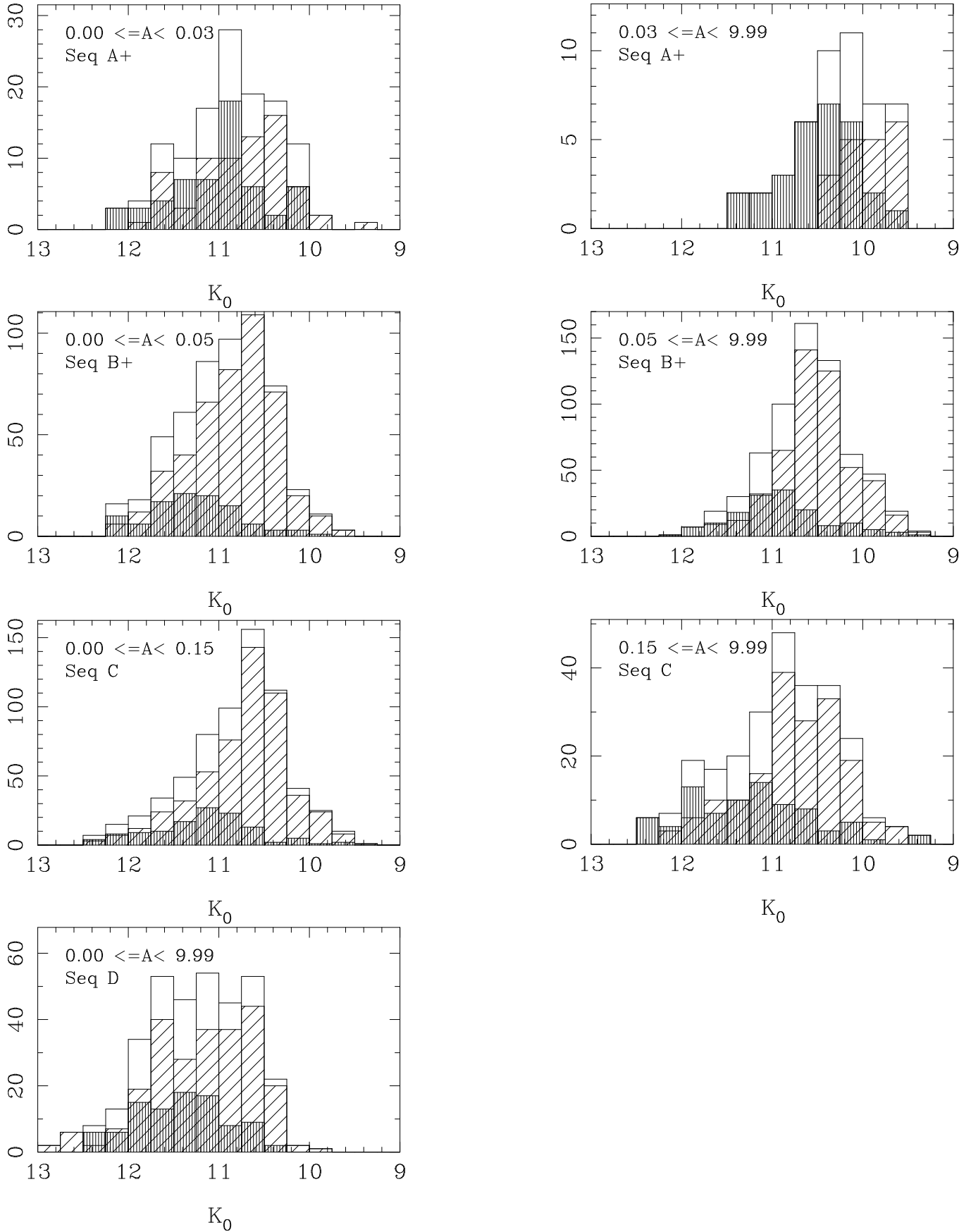

\begin{minipage}{0.42\textwidth}
\resizebox{\hsize}{!}{\includegraphics{HistoSeq_LMC_A+_0.03.ps}}
\end{minipage}
\hfill
\begin{minipage}{0.42\textwidth}
\resizebox{\hsize}{!}{\includegraphics{HistoSeq_LMC_A+_9.99.ps}}
\end{minipage}

\begin{minipage}{0.42\textwidth}
\resizebox{\hsize}{!}{\includegraphics{HistoSeq_LMC_B+_0.05.ps}}
\end{minipage}
\hfill
\begin{minipage}{0.42\textwidth}
\resizebox{\hsize}{!}{\includegraphics{HistoSeq_LMC_B+_9.99.ps}}
\end{minipage}

\begin{minipage}{0.42\textwidth}
\resizebox{\hsize}{!}{\includegraphics{HistoSeq_LMC_C_0.15.ps}}
\end{minipage}
\hfill
\begin{minipage}{0.42\textwidth}
\resizebox{\hsize}{!}{\includegraphics{HistoSeq_LMC_C_9.99.ps}}
\end{minipage}

\begin{minipage}{0.42\textwidth}
\resizebox{\hsize}{!}{\includegraphics{HistoSeq_LMC_D_9.99.ps}}
\end{minipage}
\caption[]{
Histograms of the LMC $K$-band \M\ magnitude distribution for
different ``boxes'' and $I$-band amplitude cuts. Shown are the
histograms for the M-stars (vertical lines), C-stars (hatched), and
total. The luminosity function for sequence ``D'' is likely to be
incomplete for $K$ \less 11 because of the stricter conditions to
accept periods above 800 days.
}
\label{Fig-HistoSeq-LMC}
\end{figure*}

\begin{figure*}
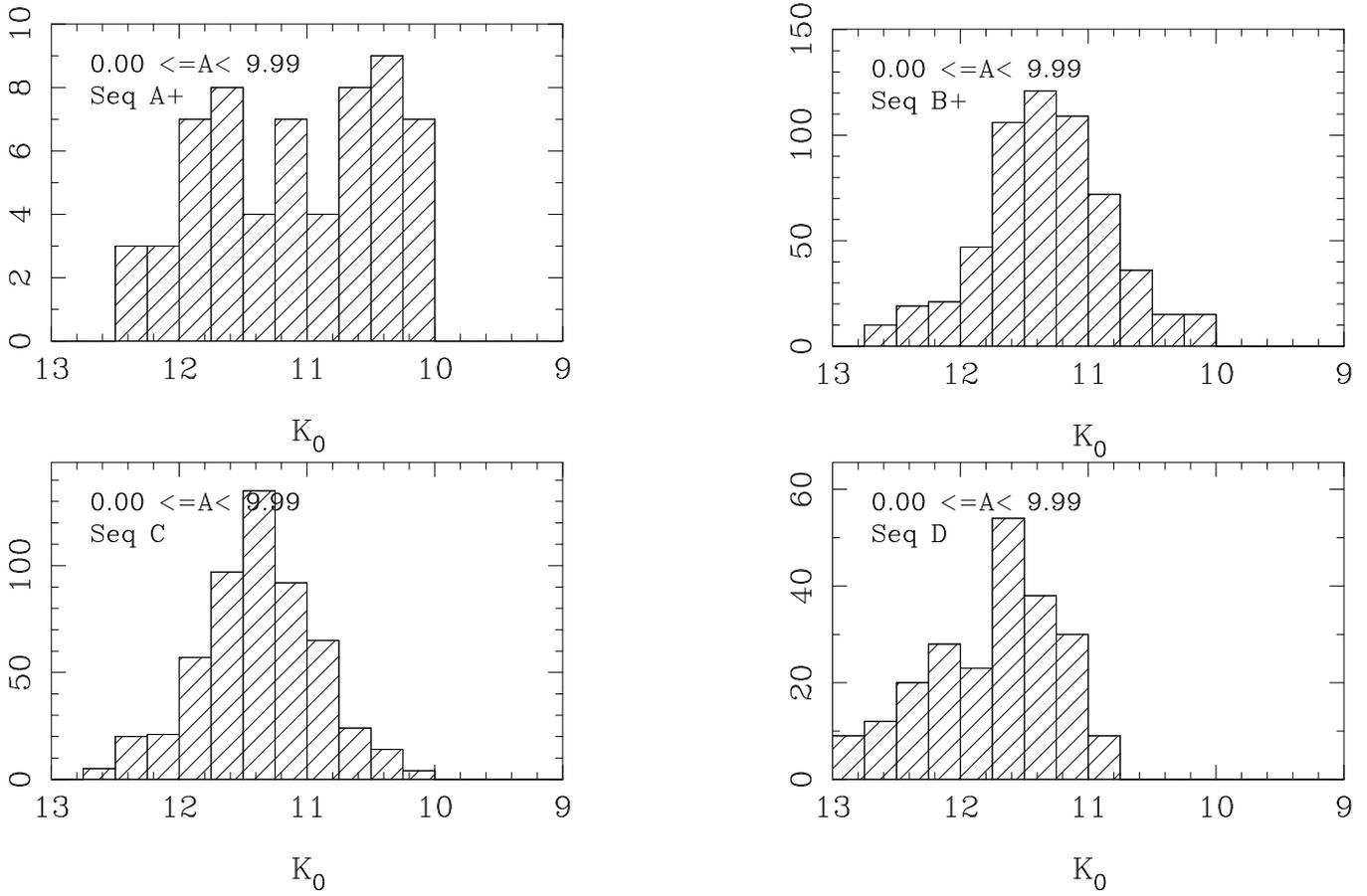

\begin{minipage}{0.42\textwidth}
\resizebox{\hsize}{!}{\includegraphics{HistoSeq_SMC_A+_9.99.ps}}
\end{minipage}
\hfill
\begin{minipage}{0.42\textwidth}
\resizebox{\hsize}{!}{\includegraphics{HistoSeq_SMC_B+_9.99.ps}}
\end{minipage}

\begin{minipage}{0.42\textwidth}
\resizebox{\hsize}{!}{\includegraphics{HistoSeq_SMC_C_9.99.ps}}
\end{minipage}
\hfill
\begin{minipage}{0.42\textwidth}
\resizebox{\hsize}{!}{\includegraphics{HistoSeq_SMC_D_9.99.ps}}
\end{minipage}
\caption[]{
As Figure~\ref{Fig-HistoSeq-LMC} for the SMC.
}
\label{Fig-HistoSeq-SMC}
\end{figure*}

Figure~\ref{Fig-Amp} shows the Amplitude-Period diagram for LMC and
SMC. They look qualitatively similar, and similar to diagrams shown in
Ita et al. (2004a) and Cioni et al. (2003). Interestingly, it appears
that for a given amplitude, the LMC M-stars have on average shorter
periods than the C-stars. This is a different manifestation of
something that could already be seen in Figure~\ref{fig:PL}, most
clearly in the panel with the cut 0.15 $\le$ Amplitude $<$ 0.45 where
sequence ``B+'' contains mainly M-stars while sequence ``C'' contains
mainly C-stars.

\begin{figure}
\begin{minipage}{0.49\textwidth}
\resizebox{\hsize}{!}{\includegraphics{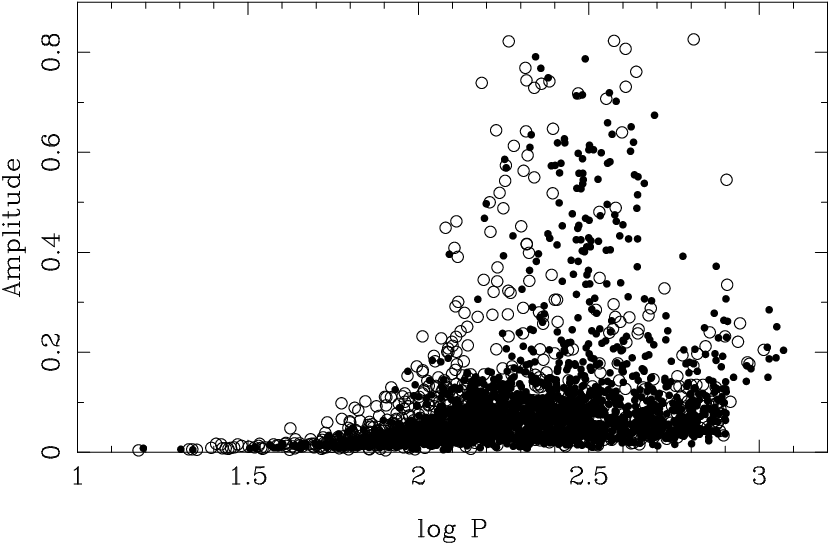}}
\end{minipage}
\hfill
\begin{minipage}{0.49\textwidth}
\resizebox{\hsize}{!}{\includegraphics{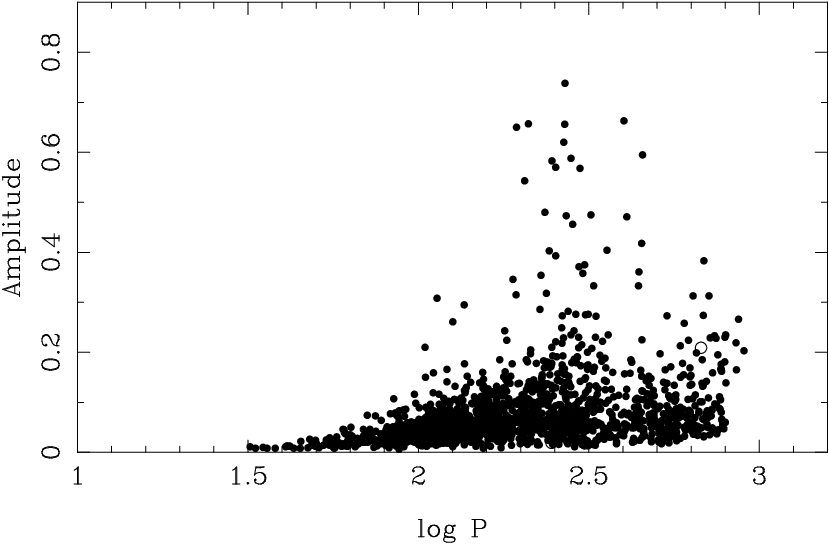}}
\end{minipage}
\caption[]{
Amplitudes versus periods for the LMC (left) and SMC
(right) variables. Symbols as in Figure~\ref{fig:PL}.
}
\label{Fig-Amp}
\end{figure}

Figure~\ref{Fig-PerCol} shows the Period-Colour diagram for LMC and
SMC for \DE\ and \M\ data. They look qualitatively similar to diagrams shown
in Lebzelter et al. (2002) and Ita et al. (2004a). Two features may be
remarked upon, (1) there are more red carbon stars in the LMC than in the
SMC (at least that are spectroscopically confirmed), (2) The M-stars
are much more spread in $(I-J)_0$ than in $(J-K)_0$ colour.

The first observation may be related to an, on average, higher mass
loss rate of AGB stars in the LMC compared to the SMC.
Section~\ref{VRO} deals in more detail with obscured objects. The
second observation is a known effect and related to the strong
temperature dependence of the molecular bands in the red part of the
optical spectrum in M-stars. Table~\ref{tab-models} lists the $(I-J)$
and $(J-K)$ colours according to the model atmospheres (for solar
metallicity) of M-stars (Fluks et al. 1994), and carbon stars (Loidl
et al. 2001, her model with a C/O ratio of 1.1). In addition, for a
few selected models the colours are listed in the last column of the
stars when obscured by 0.05 magnitudes in $K$ by dust (see Section~\ref{VRO}).

\begin{table}
\caption{Theoretical colours of M- and C-stars of solar metallicity. 
Columns 3 and 4 represent the colours for the photosphere, while, for
some models, columns 5 and 6 indicate the colours when the star is
obscured by circumstellar dust equivalent to a dimming in $K$ by 0.05 mag.}
\begin{tabular}{lrrrrr} \hline
SpT  & $T_{\rm eff}$ & $(I-J)$ & $(J-K)$ & $(I-J)$ & $(J-K)$  \\
\hline
M0  & 3850 & 1.18 & 1.08 \\
M3  & 3550 & 1.39 & 1.18 & 1.53 & 1.34 \\
M5  & 3397 & 1.86 & 1.27 \\
M7  & 3129 & 3.11 & 1.37 & 3.73 & 1.97 \\
M8  & 2890 & 3.71 & 1.37 \\
M10 & 2500 & 4.13 & 1.35 & 4.45 & 1.74 \\
C   & 3600 & 1.43 & 1.17 \\
C   & 3400 & 1.56 & 1.30 \\
C   & 3200 & 1.72 & 1.43 \\
C   & 2800 & 2.00 & 1.61 \\
C   & 2650 & 2.09 & 1.61 & 2.48 & 2.09 \\
\hline
\end{tabular}
\label{tab-models}
\end{table}

\begin{figure*}
\begin{minipage}{0.49\textwidth}
\resizebox{\hsize}{!}{\includegraphics{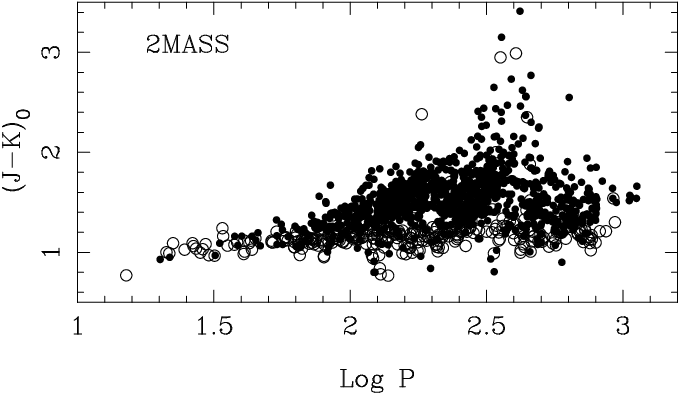}}
\end{minipage}
\hfill
\begin{minipage}{0.49\textwidth}
\resizebox{\hsize}{!}{\includegraphics{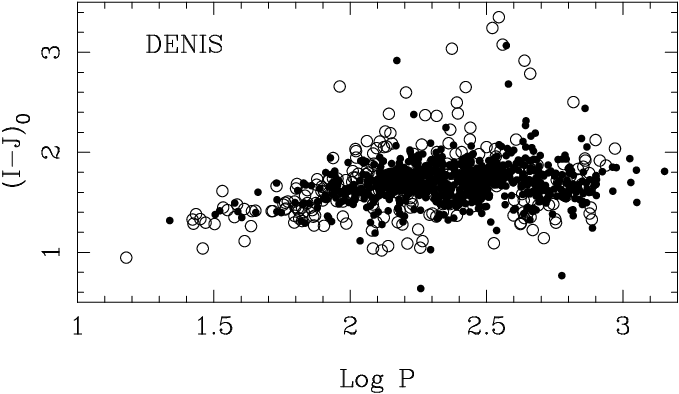}}
\end{minipage}
\begin{minipage}{0.49\textwidth}
\resizebox{\hsize}{!}{\includegraphics{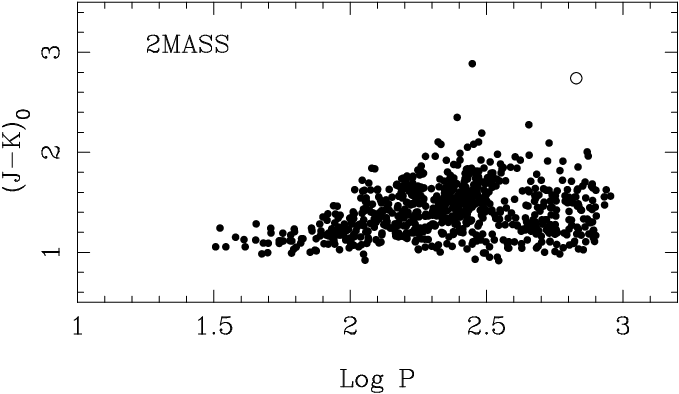}}
\end{minipage}
\hfill
\begin{minipage}{0.49\textwidth}
\resizebox{\hsize}{!}{\includegraphics{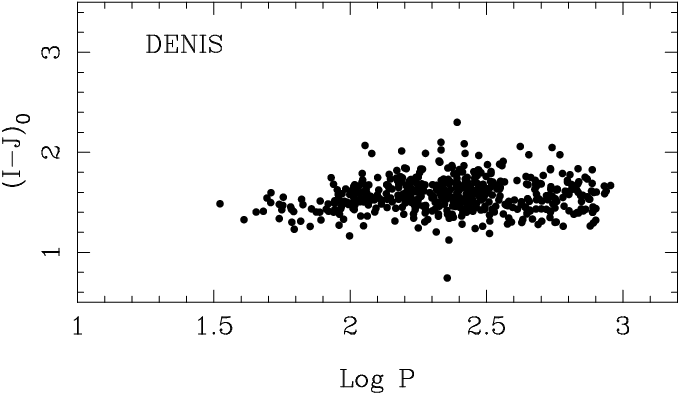}}
\end{minipage}
\caption[]{
Colour-Period diagram for the LMC (top left and top right) and SMC
(bottom left and bottom right) objects.  The period with the largest
amplitude is plotted. Symbols as in Figure~\ref{fig:PL}.
}
\label{Fig-PerCol}
\end{figure*}

\subsection{Changes in periods over time}

In this section possible changes in pulsation periods over time are
being discussed. This is possible because the periods in the Hughes
list and WBP are based on plate material typically taken between 1977
and 1984, while the MACHO, OGLE, MOA, AGAPEROS data was taken in the
late-1990s, i.e. a timespan of typically 17 years. A cross-correlation
of the \OG\ objects was performed, but this time not with
spectroscopically known objects, but rather with known LPVs with a
historical period.  Table~\ref{TAB-C} (complete table available in
electronic form) lists the results for a total of 370 stars and is
similar to Table~\ref{TAB-A}. In addition the known periods are
listed: first the historical period (Hughes list or WBP), then the
corresponding period from \OG, and when available other known periods
(with reference between parenthesis). Since multiple periods are
allowed for by the analysis of the \OG\ data, the \OG\ period quoted
is the one for which the amplitude is similar to the one corresponding
to the historical period (in some cases this then corresponds not to
the primary period as determined from the \OG\ data, but often to the
LSP). In the last column some remarks are given concerning possible
period or pulsation mode changes.

\begin{table*}
\caption[]{
First entries in the electronically available table, which list:
OGLE-field, OGLE-name, other names, periods (respectively the
historical period--in the majority of cases from the Hughes list or
WBP--then the present-day period from \OG\, and finally other available
periods with reference between parentheses), spectral type and
references and comments on period changes
}
\includegraphics[angle=+90]{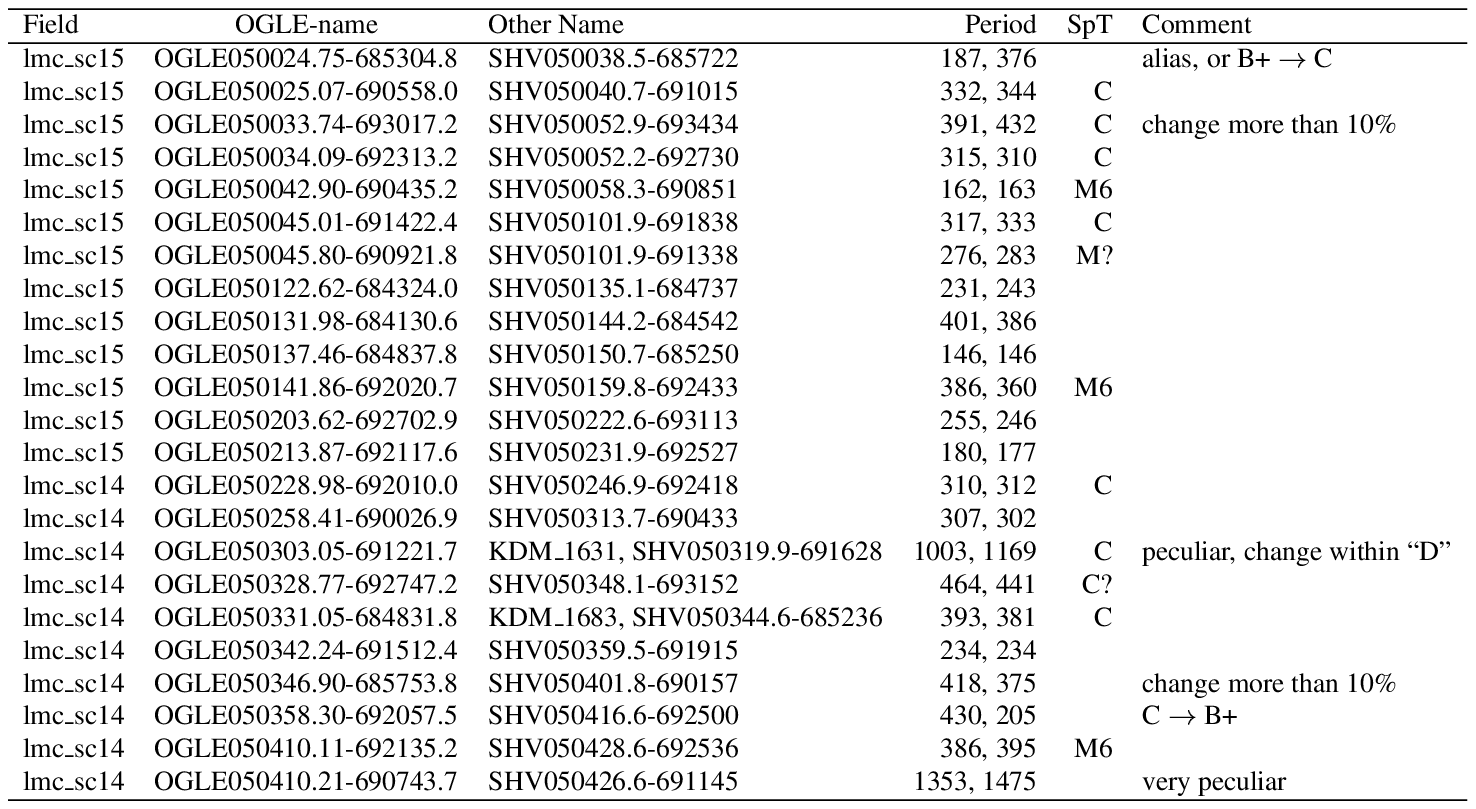}

\label{TAB-C}
\end{table*}

Of particular interest to look for period changes are the Miras that
have defined the Mira $PL$-relation over the years and that have
periods determined at multiple epochs (GLE81, WBP, FGWC, GLE03).  In
fact, GLE03 discuss 42 Miras that have defined the Mira $PL$-relation
and in particular derive periods from the MACHO database and compare
these to the historical period (FGWC, GLE81). In all but three cases
they found the periods to be constant over 2 to 3 decades. The three
are WBP-30 which changed from a 400d Mira into a small amplitude
variable with 183d period, GR0537-6740 whose period changed from 418
to 367d, and GR17 whose period changed from 780 to 729d. In the other
objects the change in period was less than 3\% of the period.

\begin{figure}
\includegraphics[width=85mm]{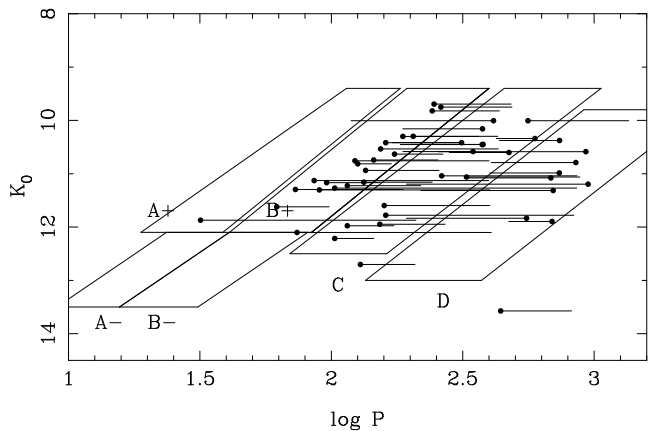}
\caption[]{
$K$-band $PL$-relation for the (LMC) stars that have undergone a
period change over the past two decades of at least 0.1 dex. Historical
and current period are connected by a line, with the current period
marked by a solid circle.
}
\label{fig-pchange}
\end{figure}

\begin{figure*}
\vskip8cm
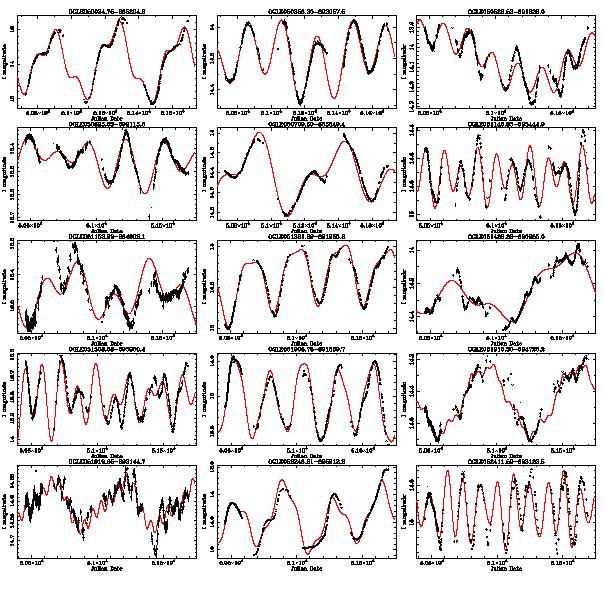
\vskip8cm
\caption[]{
Representative lightcurves of stars that have changed pulsation mode
over the last 2 decades. The fit is indicated by the (red) solid line.
Crosses indicate data points not included in the fit. Complete figure
is available in the electronic form.
}
\label{Fig-LC-LPV}
\end{figure*}

Figure~\ref{fig-pchange} shows that $K$-band $PL$-relation for the
stars that changed the period by more then 0.1 dex, and with
historical and recent period connected by a thin line. Several objects
seem to have changed period considerably and changed pulsation
mode. The data on these stars have been carefully checked in terms of
positional association, period determination, amplitude, $I$ and $JK$
magnitudes when available, etc. These stars have been marked in
Table~\ref{TAB-C}, as well as other stars that changed period by more
than 10\% (30 objects). These smaller changes may very well be real
but in most cases too small to be related to a change in pulsation
mode. Most of the changes occur from box ``C'' to ``B+'' (15 out of
36), and from ``C'' to ``D'' (8/36).  Unfortunately, the original data
points of the stars from the Hughes list are not available to directly
phase the old data points with the present day period to verify the
change in period. Figure~\ref{Fig-LC-LPV} shows some of the
lightcurves of the stars that possibly changed pulsation mode
(complete figure available in the electronic edition). As is evident
many of them are far from being regular.

Possibly the best studied case of mode switching in a Galactic
Mira-like variable is that of R Dor. Bedding et al. (1998) show that
the star switches back and forth between two pulsation period of 332
and about 175 days, on a time scale of about 1000 days. The star has
an accurate (Hipparcos) parallax and hence can be placed in
Figure~\ref{fig-pchange}.  For any reasonable distance to the LMC it
implies the star moves back and forth between sequences ``C'' and ``B+''.

\begin{figure}
\includegraphics[width=85mm]{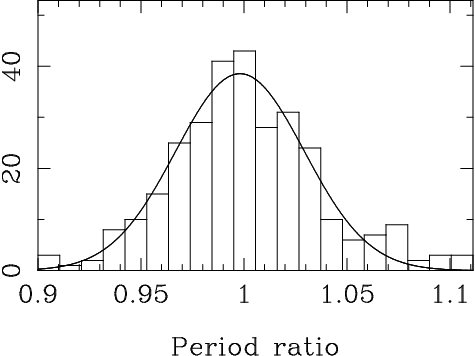}
\caption[]{
Distribution function of the ratio of the historical to the current
period near a value of unity.  The mean of all values is 1.00179. The
solid line is a Gaussian fit with a $\sigma$ of 0.031, centered at
0.9981.
}
\label{fig-pratchange}
\end{figure}

\begin{figure*}

\begin{minipage}{0.49\textwidth}
\resizebox{\hsize}{!}{\includegraphics{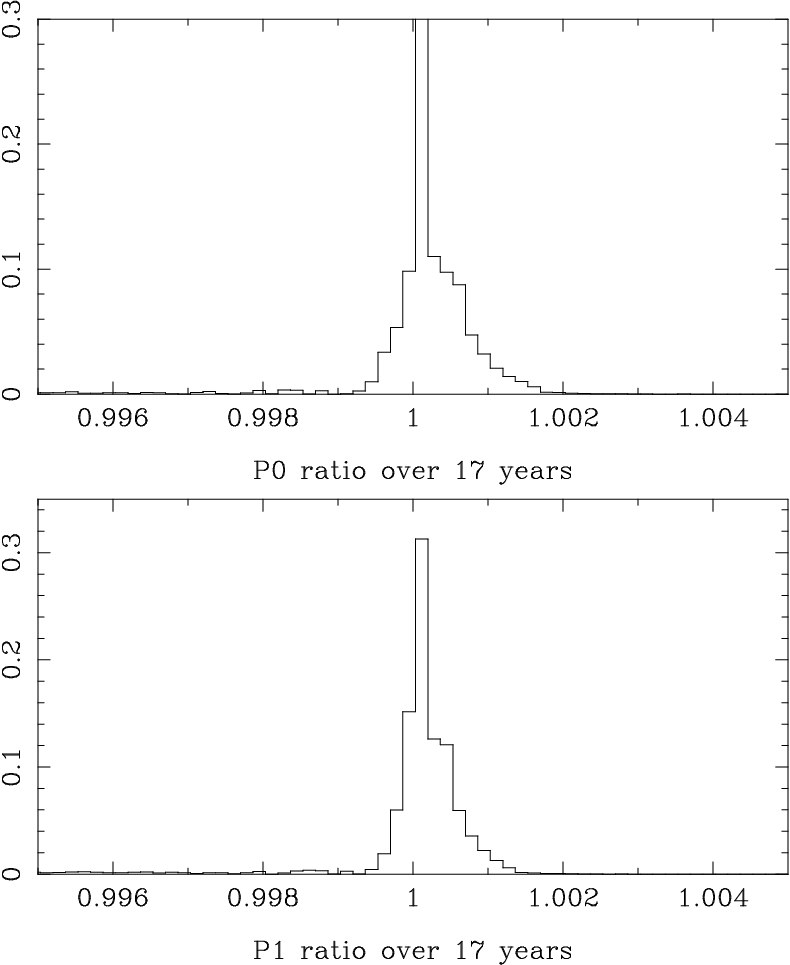}}
\end{minipage}
\hfill
\begin{minipage}{0.49\textwidth}
\resizebox{\hsize}{!}{\includegraphics{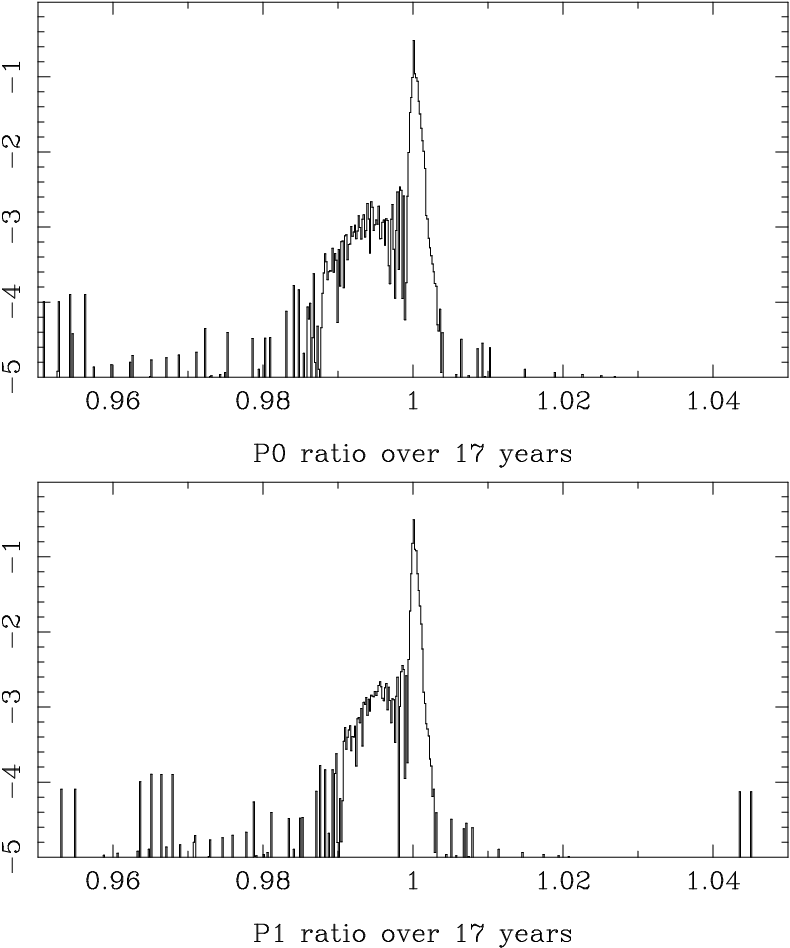}}
\end{minipage}

\caption[]{
Distribution function of the ratio of the fundamental (top panels) and
first overtone (bottom panels) pulsation period over a 17 year
timespan, on a linear (left-hand side), and logarithmic
(right-hand side) scale for a 2 \msol\ star with typical LMC
metallicity. The AGB evolution is such that on average a star evolves
to longer periods with time, but not measurably so over 17 years.
}
\label{fig-prat}
\end{figure*}

Figure~\ref{fig-pratchange} shows the distribution of the ratio
(current period / historical period) near a value of 1. The observed
mean is 1.00018. Also a Gaussian fit to this distribution is shown to
guide the eye, which has a dispersion of 0.031. As stars evolve to
cooler temperature and lower mass on the AGB one might expect an
increase of the pulsation period. To verify this theoretically, the
synthetic AGB evolution code based on the formulation in Wagenhuber \&
Groenewegen (1998) was used for a typical star with 2 \msol\ initial
mass and metallicity $Z$ = 0.008. The model takes into account the
changes in stellar parameters over the thermal pulse cycles, and
assumes that pulsation takes place over the entire TP-AGB, i.e. no
cessation of pulsation or mode switches during any phase of the stars'
TP-AGB evolution. For every time step the fundamental and first
overtone pulsation period was calculated from $P_0 = 0.00851 R^{1.94}
M^{-0.90}$ and $P_1 = 0.038 R^{1.5} M^{-0.5}$ (Wood 1990; $R$ being
the stellar radius in solar radii, and $M$ the total mass in solar
masses).  Figure~\ref{fig-prat} shows the distribution function of
$P_0$ and $P_1$ over a timespan of 17 years for such an object, both
on a linear and logarithmic scale. The AGB evolution is such that on
average a star evolves to longer period with time, but almost not
measurably so over 17 years. Indeed the averages are 1.00024 for
fundamental mode and 1.00015 for first overtone. This implies that for
any individual star one must be able to determine periods to fractions
of a day to be able to detect period changes due to evolution, which
seems unrealistic since the lightcurves are rather complicated and not
mono-periodic in the majority of cases. The fact that the observed
mean of the period ratios over a 17 year timespan is close to the
predicted one probably indicates good fortune rather than to indicate
that, in a statistical sense, the predicted evolutionary effect has
indeed been detected. The width of the observed distribution is much
wider than predicted by the models, and is also wider than expected
from the errors in the observed period determinations alone. The
median period of the stars listed in Table~\ref{TAB-C} is about 290
days, and in the overall majority of cases such a period has been
determined with an accuracy of 0.7 days, or better. This implies one
would expect ratios near unity with an error of about 0.005 or
smaller. This would suggest that the width of the distribution is
real.

In addition, there exist a few examples of LPVs whose period {\it
decrease} over a longer timescale (Wood \& Zarro 1981). Zijlstra et
al. (2002) show that for the Mira R Hya $\dot{P}$ is about $-1.6\;
10^{-3}$ between 1770 and 1950 AD. Although its period has actually
stabilised at 385d since then, let us assume it would follow its
historical trend. Over a 17 year period it would decrease its period
by 9.9 days and would have a period ratio, as defined and used in
Figures~\ref{fig-pratchange} and \ref{fig-prat}, of 0.975. These
values are predicted by the stellar evolutionary calculations
mentioned above (and by the calculations in Wood \& Zarro) but with a
very small probability.  The fact that the observed distribution in
Figure~\ref{fig-pratchange} is wider than expected, based on the
errors alone, suggests that other phenomena than the ``global''
increase in period over an AGB stars lifetime dominate this
distribution, or that a model assumption is incorrect. Either
pulsation does not take place at all phases of AGB evolution which
would influence the theoretically predicted distribution function, or
other physical phenomena play a role. Zijlstra et al. (2002) mention
and refer to weak chaotic behaviour and the effects of the
non-linearity of the pulsation in the case of R Hya.

\subsection{Very red objects}
\label{VRO}

\begin{figure*}
\begin{minipage}{0.49\textwidth}
\resizebox{\hsize}{!}{\includegraphics{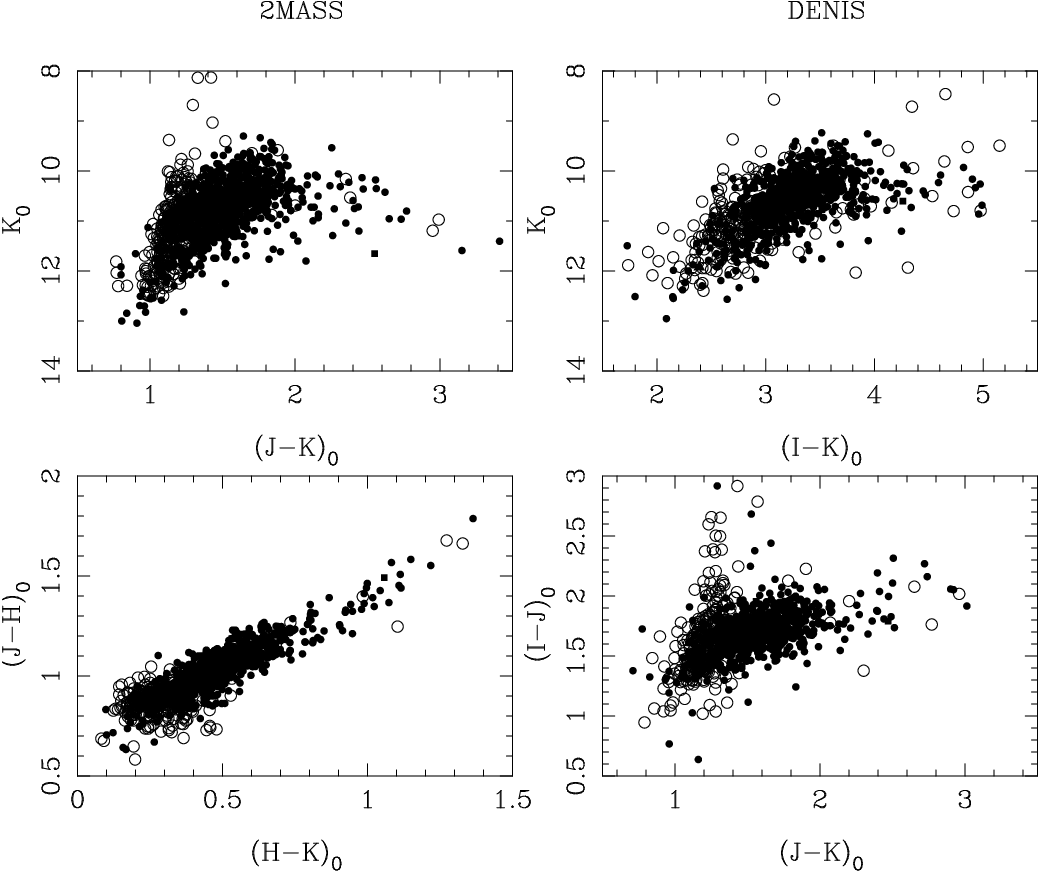}}
\end{minipage}
\hfill
\begin{minipage}{0.49\textwidth}
\resizebox{\hsize}{!}{\includegraphics{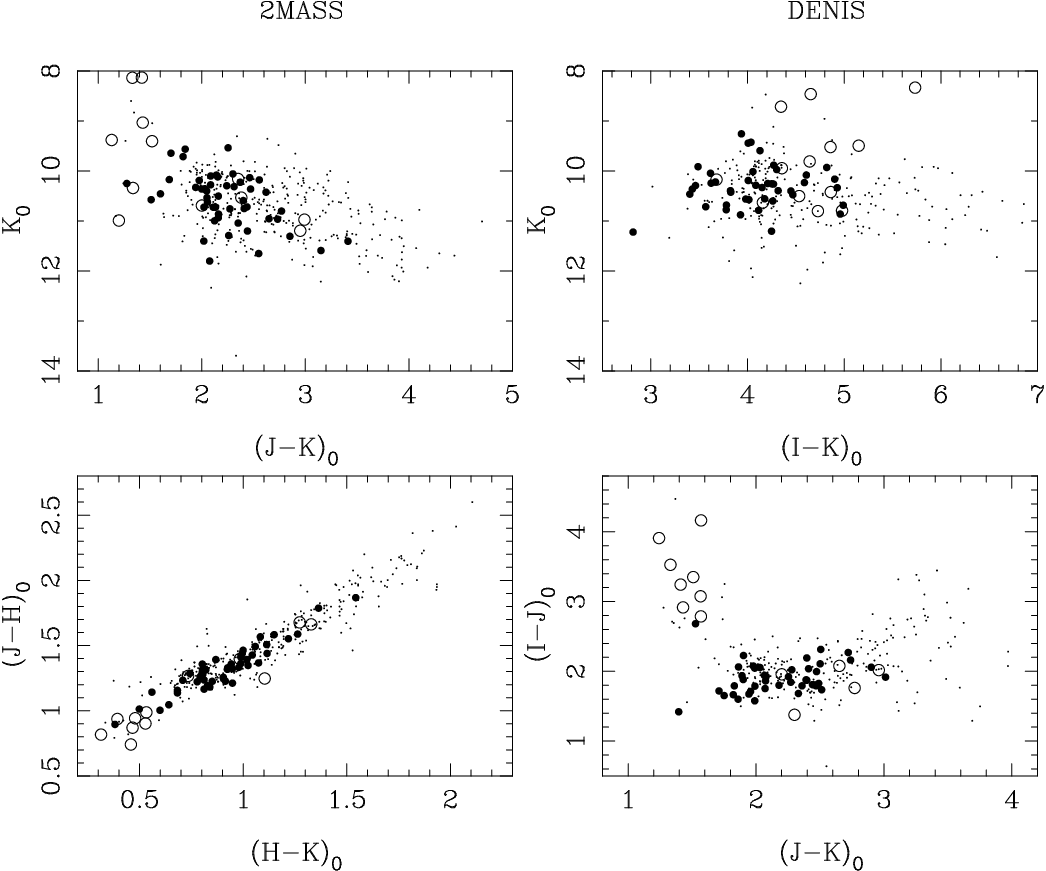}}
\end{minipage}

\begin{minipage}{0.49\textwidth}
\resizebox{\hsize}{!}{\includegraphics{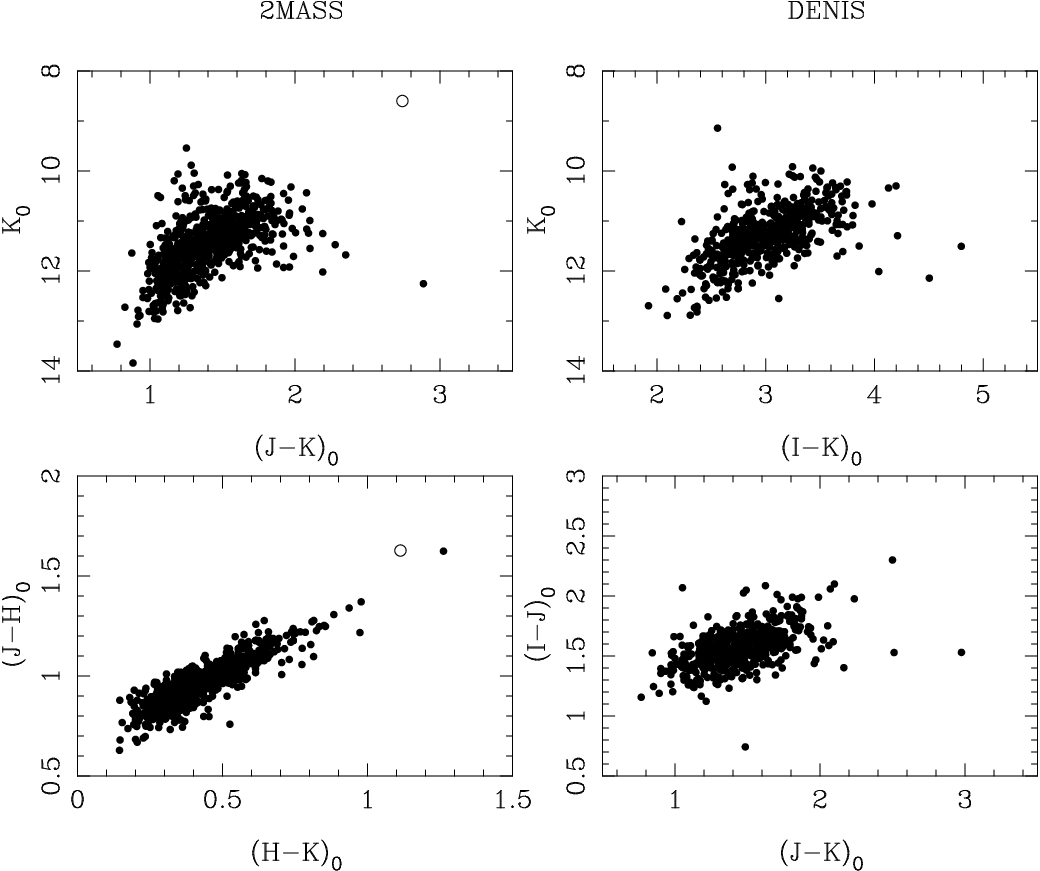}}
\end{minipage}
\hfill
\begin{minipage}{0.49\textwidth}
\resizebox{\hsize}{!}{\includegraphics{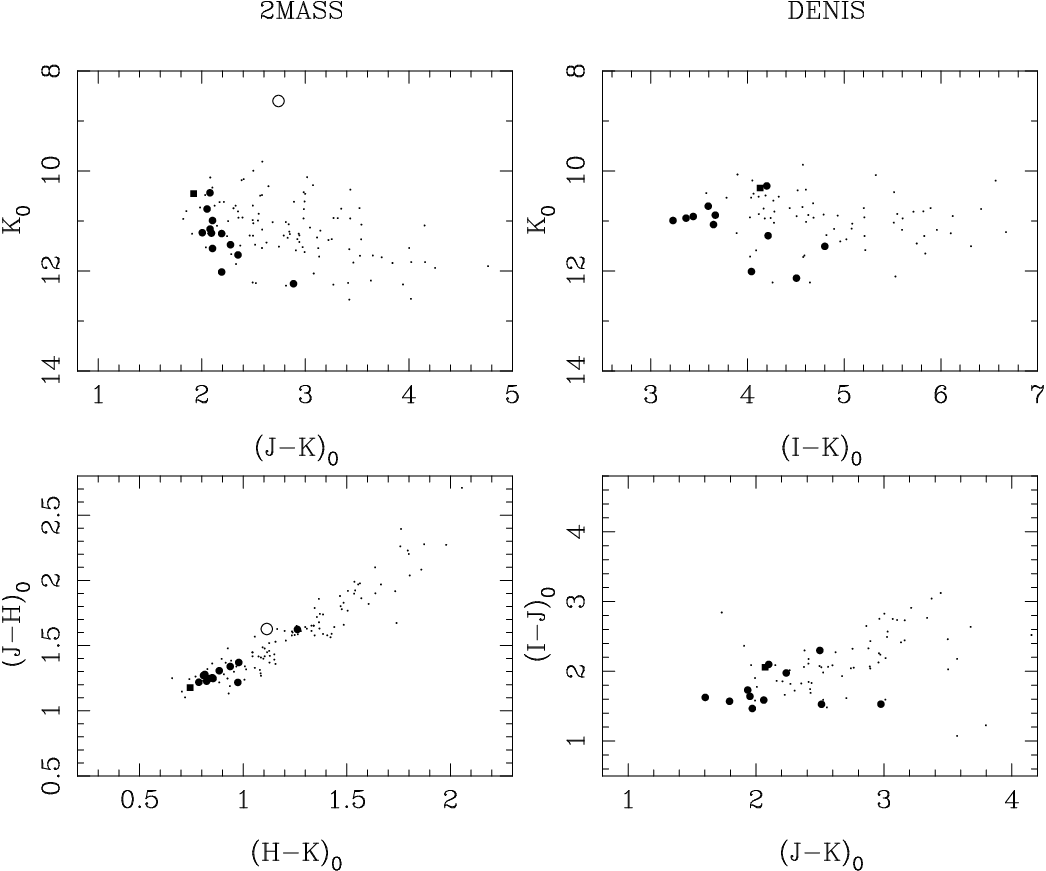}}
\end{minipage}
\caption[]{
LMC (top 2 rows) and SMC (bottom 2 rows) colour-colour diagrams using
\M\ and \DE\ photometry for the spectroscopically selected sample
(left 2 columns) and the infrared selected sample (right 2
columns). Note the difference in scale!  Symbols as in
Figure~\ref{fig:PL}, with dots indicating objects without
spectroscopic classification.
}
\label{Fig-CC}
\end{figure*}

Since the spectroscopically selected sample, for given luminosity,
will be biased against stars with heavy mass loss, and hence fainter
$I$ magnitudes, the present section discusses an infrared selected
sample with red infrared colours, which will complement the sample of
spectroscopically selected AGB stars. Figure~\ref{Fig-CC} shows the
\DE\ $K_0 - (I-K)_0$ magnitude-colour, $(I-J)_0 - (J-K)_0$
colour-colour, and the \M\ $K_0 - (J-K)_0$ magnitude-colour, $(J-H)_0
- (H-K)_0$ colour-colour diagrams for both LMC and SMC for the
spectroscopically selected sample in the two leftmost columns. Based
on this diagram it was decided to investigate the pulsation
characteristics of objects that have \DE\ $(I-K)_0 > 4.0$ and
$({\sigma}_{\rm I}^2 + {\sigma}_{\rm K}^2) / (I-K)_0 < 0.01$ or \M\
$(J-K)_0 > 2.0$ and $({\sigma}_{\rm J}^2 + {\sigma}_{\rm K}^2) /
(J-K)_0 < 0.01$.

\begin{figure*}
\vskip8cm
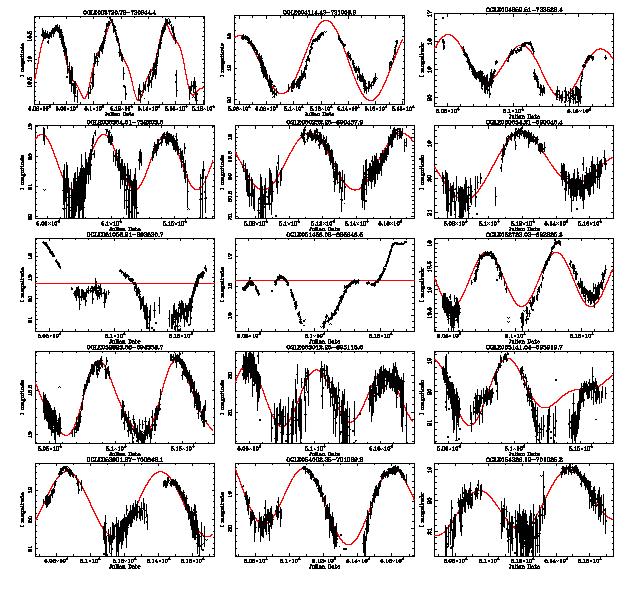
\vskip7cm
\caption[]{
Lightcurves of the reddest stars in $(J-K)$. Note the faintness in
$I$. Nine of the 15 stars have been detected in the MSX survey.
Complete figure is available in electronic form.
}
\label{Fig-LC-IR}
\end{figure*}

The infrared selected sample contains 577 objects (137 SMC, 442 LMC)
that fullfill these limits and Figure~\ref{Fig-LC-IR} shows the
lightcurves of the reddest stars in $(J-K)$ among them (the full
figure is available in the electronic edition). Tables~\ref{TAB-E} and
\ref{TAB-F} show the first entries with information similar to
Tables~\ref{TAB-A} and \ref{TAB-B}. The colour-colour diagrams are
shown in the two right side columns in Figure~\ref{Fig-CC}. Some stars
from the spectroscopically selected sample also appear in the infrared
selected sample, mostly M-stars for which it is was shown before that
the later spectral types are reasonably red in $(I-K)$, certainly
compared to carbon stars for a given $(J-K)$.

Figure~\ref{fig:PLIR} finally shows the $K$-band $PL$-relation for the
large amplitude variables in the infrared selected sample. Compared to
stars that are spectroscopically known they appear to be fainter at a
given period. This must be largely due to the dust obscuration. An
identical effect was shown by Wood (2003) who specifically looked at
MSX sources.  To illustrate this, Figure~\ref{FIG-SED} shows the
spectral energy distribution of one of the stars in
Figure~\ref{fig:PLIR}, namely OGLE050854.21-690046.4 or MSX 83. For a
distance of 50.1 kpc, a luminosity of 22100 \lsol\ is derived and a
mass loss rate of 7.6 $\times 10^{-6}$ \msolyr.

\begin{table*}
\caption[]{
The infrared selected sample. As Table~\ref{TAB-A}.
}
\includegraphics[angle=+90, width=190mm]{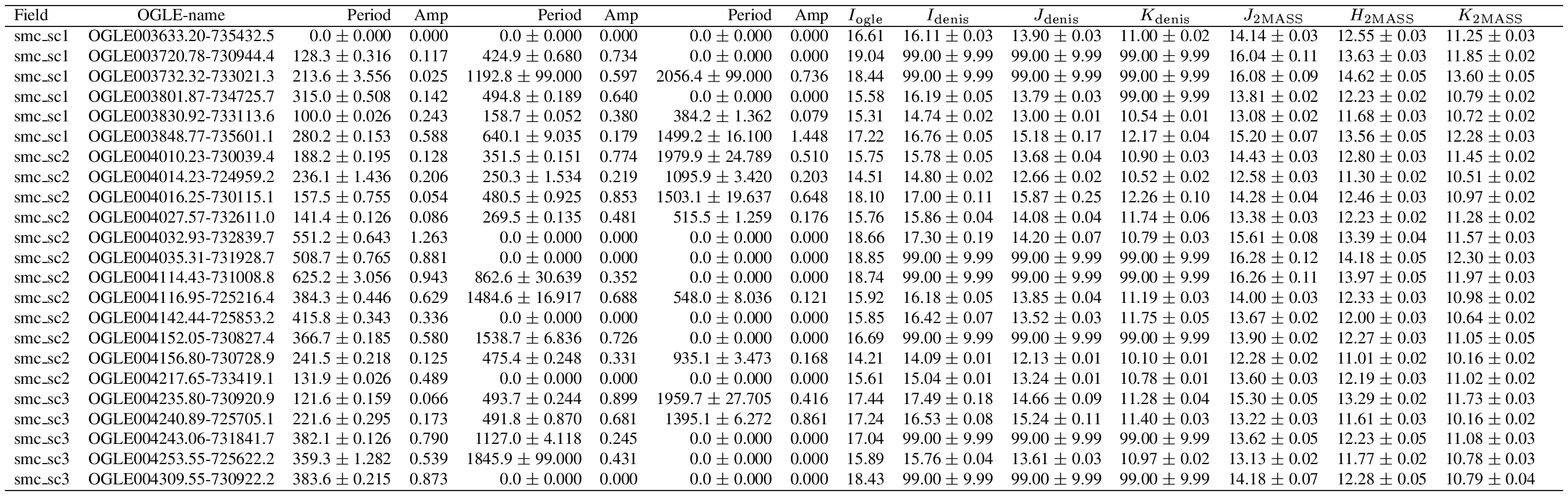}
\label{TAB-E}
\end{table*}

\begin{table}
\caption[]{
The infrared selected sample. As Table~\ref{TAB-B}. 
}
\includegraphics[angle=+90, width= 190mm]{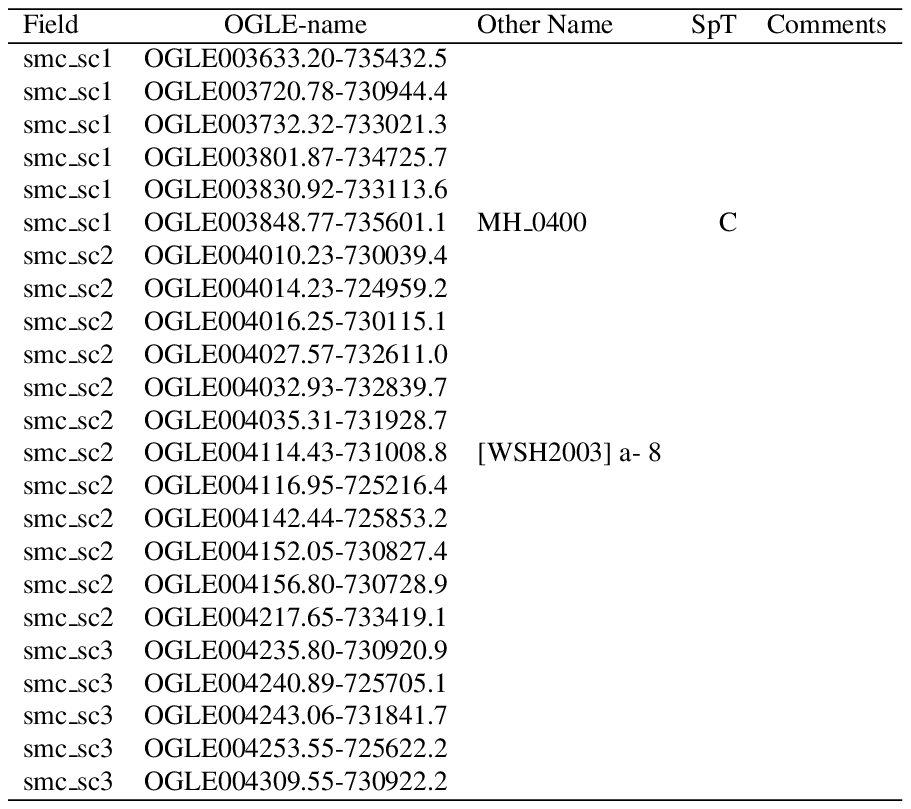}
\label{TAB-F}
\end{table}

\begin{figure*}[ht]
\begin{minipage}{0.48\textwidth}
\resizebox{\hsize}{!}{\includegraphics{K-P_SMC_0.45_9.99_IR.ps}}
\end{minipage}
\hfill
\begin{minipage}{0.48\textwidth}
\resizebox{\hsize}{!}{\includegraphics{K-P_LMC_0.45_9.99_IR.ps}}
\end{minipage}
\caption{
$K$-band $PL$-relation for the large amplitude variables in the 
IR selected sample, for the SMC (left) and LMC (right). 
}
\label{fig:PLIR}
\end{figure*}

\begin{figure}
\includegraphics[angle=-90, width=85mm]{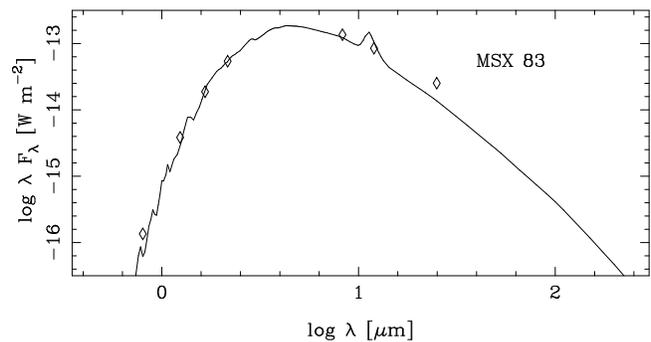}
\caption[]{
SED of one of the stars in the IR selected sample, MSX 83 annex
OGLE050854.21-690046.4. Plotted are the \OG\ $I$, \M\ $JHK$, MSX $A$
band and IRAS 12 and 25 $\mu$m fluxes. A carbon star central star
model is assumed.
}
\label{FIG-SED}
\end{figure}

\section{Summary and discussion}

This paper addresses several aspects of the pulsational character of
late-type stars in the Magellanic Clouds. The main focus is on the
$K$-band $PL$-relation of almost 2300 spectroscopically confirmed
M-,S- and C-stars. This sample avoids to make the clearly incorrect
approximation made in other studies that M-stars and carbon stars can
be separated at a colour $(J-K)$ = 1.4.

The present observations however do not allow the presentation a
comprehensive picture of the evolution of pulsation periods of M- and
C-stars. This would require a more detailed (AGB star) population
synthesis study including pulsation properties. As previous MCs
studies have found for AGB stars in general (Wood et al. (1999), Wood
(2000), Noda et al. (2002), Lebzelter et al. (2002), Cioni et
al. (2003), Ita et al. (2004a,b) and Kiss \& Bedding (2003, 2004)), it
is found specifically that both M-, and C-stars tend to occupy
preferentially sequences B+ and then C for increasing amplitude.  For
a given amplitude, the C-stars tend to have the longer period, and,
for a every sequence they are more luminous. This effect has
previously been observed in MC clusters (e.g. FMB) and is in
qualitative agreement with evolutionary calculations that predict that
C-stars evolve from M-stars.

Many objects have one period that falls in box ``D''. In the
spectroscopically selected sample, 211 of 859 SMC stars (= 24.6\%)
have a period that falls in box ``D'', and 318 of the 1418 LMC stars
(= 22.4\%, namely 229/1064 = 21.5\% of C-stars and 89/354 = 25.1\% of
M-stars). As will be discussed below, for at least some stars of the
IR selected sample this is due to the fact that they are weaker in $K$
because of dust obscuration. For the overall majority of the stars in
the spectroscopically selected sample this is not an issue. The reason
why some late-type stars appear on that location of the $PL$-diagram
is unexplained, see the discussion in Olivier \& Wood (2003), Wood
(2003, 2004). The classical large-amplitude Mira variables appear on
sequence ``C'' (see the last panels in Figure~\ref{fig:PL}) and are
believed to be fundamental-mode pulsators, hence longer (radial mode
pulsation) periods should not exist.  The present paper does not shed
light into the nature of the LSP phenomenon, except that
Figure~\ref{Fig-HistoSeq-LMC} indicates that the $K$-band luminosity
function for sequence ``D'' is essentially the same for the M- and
C-stars, while for sequences ``A,B,C'' the C-stars are brighter, as
expected from an evolutionary point of view. The luminosity
function of objects on sequence ``D'' is fainter from those of the
other sequences. This is due to the fact that periods longer than 800
days are underrepresented because of stricter selection rules. This
affects predominatly sequence ``D'' objects with $K_0$ \less 10.5
mag. The fact that the fraction of all object with periods on
sequence ``D'' is essentially the same for C- and M-stars and that the
$K$-band luminosity of C- and M-stars of sequence ``D'' objects is
very similar suggests that the LSP phenomenon is unrelated to chemical
type, and hence seems unrelated to a pulsation phenomenon.

For a few hundred variables it was possible to look for period changes
over a timespan of typically 17 years. Almost all come from the
studies by Hughes (1989) and Hughes \& Wood (1990). They identified
medium to large amplitude variables from photographic material using
typically 21 observations in the time span 1977 to 1984. Out of 370
objects, 36 have been identified that seemingly changed pulsation mode
(or at least changed ``box'') between $\sim$1980 and the time of the
\OG\ observations, and another 30 objects that changed pulsation
period by more than 10\%. This ratio of about 10\% (36/370) is similar
to the study of GLE03 who found large period changes in 3 out of 42
Mira variables studied. A caveat is that the original historical data
points have never been published, and it would certainly be preferable
to be able to phase the old data with the current period to see if in
fact the period change is real. For the moment I consider the 10\%
change of pulsation sequence over $\sim$17 years as an upper limit. By
comparison, Zijlstra \& Bedding (2003) find that only of order of 1\%
of well-known Miras show evidence for period changes. The
understanding for MC objects may improve because of ongoing (e.g. {\sc
ogle-iii}) and future surveys. 

Finally a sample of stars was studied selected on infrared colours,
namely redder than the majority of the spectroscopically selected
sample. It should be pointed out that there is no proof that these are
AGB stars (except for the few ones in overlap with the
spectroscopically selected sample) and they make suitable targets for
spectroscopic follow-up to determine their spectral type. Many of
these stars also have a period located in box ``D'' but in this case
the effect of obscuration by dust must be considered. This also has
implications when using samples of variables to determine $K$-band
$PL$-relations.  In the last section one explicit example was shown,
namely MSX 83, for which the SED was constructed and fitted with a
dust radiative transfer model. Its period of 611 day, and $K$
magnitude of 10.58 would put it in ``box D''. However, when running
the radiative transfer model without mass-loss the $K$-magnitude
brightens to 9.17 mag (this value is somewhat dependent on the central
star model atmosphere assumed) putting it on the extension of box
``C'', consistent with the expected location for a star with an
pulsation amplitude of 0.66 mag.

In reverse this implies that when studying the $K$-band $PL$-relation,
and when multi-colour data is available, a cut-off in colour should be
applied in order to avoid a bias by including stars that are dust
obscured in $K$.  Although this seems obvious, a quantification of
where this cut-off should be placed and its actual application are
rare in the literature; In fact I could only find one instance. Glass
et al. (1995) mention they exclude some faint outliers with $(K-L)
\sim 2$ in the Sgr {\sc i} bulge field (which corresponds to roughly
$(J-K) \sim$ 3.5, Glass, 1986), but did not impose a colour criterium
a-priori. In other papers where dust obscuration in the studied
variables should play a role the bolometric $PL$-relation is studied
(e.g. WFLZ), which circumvents this problem in a natural way.

It is difficult to give an exact colour cut-off to apply, since this
depends on the colours involved, the dust properties and the evolution
of mass-loss on the AGB. Based on the calculations presented in the
last columns in Table~\ref{tab-models} stars with colours $(J-K)_0 >
2.0$ should certainly be avoided, and a stricter criterium would
be to include M-stars only when $(J-K)_0 <$ 1.4 and C-stars only when
$(J-K)_0 <$ 1.7.

\begin{figure*}[ht]
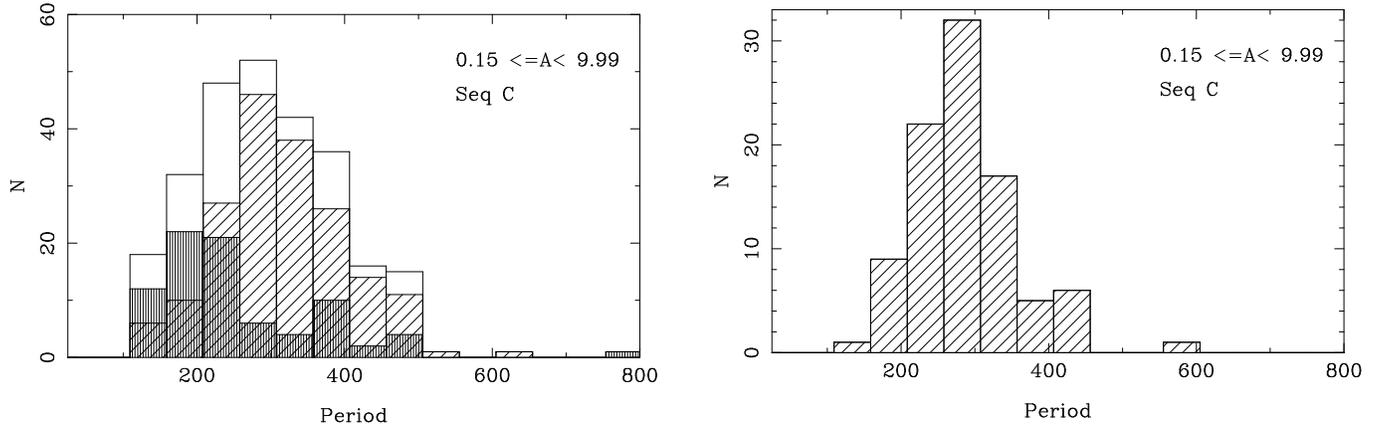

\begin{minipage}{0.48\textwidth}
\resizebox{\hsize}{!}{\includegraphics{PerDistLmc_C_0.15_9.99.ps}}
\end{minipage}
\hfill
\begin{minipage}{0.48\textwidth}
\resizebox{\hsize}{!}{\includegraphics{PerDistSmc_C_0.15_9.99.ps}}
\end{minipage}
\caption{
Period distribution of the LMC (left panel) and SMC (right panel)
variables in box ``C'' with $I$-band amplitudes larger than 0.15
magnitudes. Shown are the histograms for the M-stars (vertical lines),
C-stars (hatched), and total.
}
\label{PerDist}
\end{figure*}

Finally, applying these latter cut-offs to the spectroscopically
selected stars in box ``C'' with amplitudes $>0.45$ mag and $>0.15$
mag, to improve the statistics, the $K$-band $PL$-relations listed in
Table~\ref{tab-pl} have been derived, where also some relations from
the literature are listed. The period distribution of these
samples is shown in Figure~\ref{PerDist}. It would be interesting to
fit the SEDs of all infrared selected stars to be able to include the
dust-correced $K$-magnitudes in these $PL$-relations.

There is very good agreement between the new relations (including what
appears to be the first Mira $K$-band $PL$-relation in the SMC) and
previous works. The formal error on the zero point and slope have
become smaller, because of the larger sample size, but the overall rms
are still larger because the photometry used in FGWC and GLE03 are
averages over the lightcurve while the present data are at best
averages of two measurements.

\begin{table*}
\caption{$K$-band $PL$-relations in box ``C'', of the form $m_{\rm K}= a \log P + b$ }
\begin{tabular}{ccrcl} \hline
       $a$       &       $b$        &  N  &  rms & remarks  \\
\hline
$-3.78 \pm 0.24$ & $20.17 \pm 0.58$ &  44 & 0.26 & O-stars, LMC, Amplitude $> 0.45$ mag \\
$-3.50 \pm 0.27$ & $19.42 \pm 0.67$ &  58 & 0.22 & C-stars, LMC, Amplitude $> 0.45$ mag \\
$-4.14 \pm 0.85$ & $21.34 \pm 2.06$ &  14 & 0.25 & C-stars, SMC  Amplitude $> 0.45$ mag \\
$-3.52 \pm 0.16$ & $19.56 \pm 0.38$ &  83 & 0.26 & O-stars, LMC, Amplitude $> 0.15$ mag \\
$-3.56 \pm 0.15$ & $19.57 \pm 0.38$ & 181 & 0.25 & C-stars, LMC, Amplitude $> 0.15$ mag \\
$-3.67 \pm 0.24$ & $20.22 \pm 0.60$ &  92 & 0.24 & C-stars, SMC, Amplitude $> 0.15$ mag \\
 \\
$-3.47 \pm 0.19$ & $19.48 \pm 0.45$ &  29 & 0.13 & M-Miras, LMC, FGWC \\
$-3.52 \pm 0.21$ & $19.64 \pm 0.49$ &  26 & 0.13 & M-Miras, LMC, GLE03 \\
$-3.30 \pm 0.40$ & $18.98 \pm 0.98$ &  20 & 0.18 & C-Miras, LMC, FGWC \\
$-3.56 \pm 0.17$ & $19.64 \pm 0.42$ &  54 & 0.25 & C-Miras, LMC, Groenewegen \& Whitelock (1996)  \\
\hline
\end{tabular}
\label{tab-pl}
\end{table*}

\acknowledgements{ 

The author would like to thank Laurent Eyer for interesting
discussions, Peter Wood (MSSSO), Mathias Schultheis (IAP) and
Maria-Rosa Cioni (ESO) for providing computer readable versions of
relevant data, Yoshifusa Ita for providing the boundaries of the
sequences in the $K$-band $PL$-diagram in Ita et al. (2004a), Joris
Blommaert for commenting on an earlier draft, and an anonymous referee
for a very carefull reading of the manuscript.

This research has made use of the SIMBAD database, operated at CDS,
Strasbourg, France.

This publication makes use of data products from the Two Micron All
Sky Survey, which is a joint project of the University of
Massachusetts and the Infrared Processing and Analysis
Center/California Institute of Technology, funded by the National
Aeronautics and Space Administration and the National Science Foundation.

This paper utilizes public domain data originally obtained by the
MACHO Project, whose work was performed under the joint auspices of
the U.S. Department of Energy, National Security Administration by the
University of California, Lawrence Livermore National Laboratory under
contract No. W-7405-Eng-48, the National Science Foundation through
the Center for Particle Astrophysics of the University of California
under cooperative agreement AST-8809616, and the Mount Stromlo and
Siding Spring Observatory, part of the Australian National University

}

{}

\appendix

\section{Step~1: Fourier analysis and light curve fitting}

At the heart of the first code are the {\sc numerical recipes}
subroutines {\it fasper} and {\it mrqmin} that perform a Fourier
transformation and a (weighted) linear least-squares fitting, respectively. The
function that is fitted to the data has the following form:
\begin{equation}
I(t) = I_0 + \sum_{i=1}^{i=n_{\rm max}} 
\left( A_{\rm i} \sin (2 \pi \; t \; e^{f_{\rm i}}) + 
       B_{\rm i} \cos (2 \pi \; t \; e^{f_{\rm i}}) \right)
\end{equation}
This is a more suitable form for fitting purposes than the equivalent
form (with $\omega = e^f$):
\begin{displaymath}
I(t) = I_0 + \sum_{i=1}^{i=n_{\rm max}} 
C_{\rm i} \sin (2 \pi \; t \; {\omega_{\rm i}} + \phi_{\rm i} )
\end{displaymath}
Up to three periods are fitted ($n_{\rm max}$ = 3), following Wood et
al. (1999).  These two subroutines are called alternatingly,
subtracting the best fit sofar from the data, and then performing a
Fourier analysis on the residual. If a significant peak is found, a
fit including the next term in the series in Eq.~1 is included. This
is repeated until no significant peak in the power spectrum is found
or the third period has been fitted.

The Fourier analysis is done with the subroutine {\it fasper}. Inputs
to it are the time and magnitude arrays. In addition one has to
specify two parameters, {\it ofac} and {\it hifac}, that indicate a
``typical oversampling factor'' and the maximum frequency in terms of
a ``typical'' Nyquist frequency.

In the present work {\it ofac } = 22 and {\it hifac} = 0.8 are used.
The latter parameter is the determining factor in both the
computational speed, and the shortest period that can be found. For
example, some tests to correctly identify the main period in an RR
Lyrae object with a known period of 0.55 days required {\it hifac = 10.0}. 
In this configuration the code would be more than a factor of 10 slower. 
At the same time it implies that in the configuration used in the
present paper there is a bias in the detection of periods shorter than
about 6 days, of no consequence as the focus is on AGB stars.

The output of {\it fasper} are the frequency where the peak occurs and
a number indicating a {\it significance}. One of the main parameters
in the code is to provide the critical cut-off above which a period is
not considered to be significant. In the present work 
{\it significance = $ 5.5 \times 10^{-11}$ } is used, and this was
determined empirically, by visually inspecting many lightcurves.

The code can be run in a single-star mode (for fine tuning) or in an
automatic mode. It should be pointed out that some of the features and
parameters just described and that will be described below have been
determined only empirically. In fact, the cycle of Steps~1 and 2, and
partly Step~3, has been repeated a few times to check the various
steps in detail and come to the final choice of the parameters. The
content of the code is now described in detail.

\begin{itemize}

\item Read in the files with \M\ and \DE\ objects that are within
3\arcsec\ of \OG\ objects. 

Read in the file which contains the path names to the 68~000 files
which contain the $I$-band data. 

The steps below are either done for one star (in single-star mode) or
for all.

\item Read in the path-name (e.g. smc\_sc5/OGLE005013.32-731112.6.mag),
and construct the right ascension and declination from it. Read in the
individual Julian Dates, $I$-band magnitudes and errors. If there are
less than 11 data points, exit. Otherwise enter the light curve fitting
subroutine.

\item Loop over the data points, and remove those with error $\mbox{``$-99$''}$.
Loop over the remaining data points and determine the largest time gap
between two subsequent data points in time. If there are less than 30
data points or a time gap larger than 250 days, exit.

\item Order the data points according to the error bar in the
$I$-magnitude.  Automatically reject 5\% of the data with the largest
errors. Order the data points according to the
$I$-magnitude. Automatically reject the 3 brightest data points.  This
last step was introduced when it became clear that a few very bright
(spurious ?) outliers in faint sources severely biased the light curve
fitting because these outlying points carry so much weight. In bright
sources this procedure has no bearing on the outcome. The final data
arrays with time, $I$-magnitude and error in $I$-magnitude are
constructed, and written to file, so that they can be read in by an
independent time-analysis code ({\sc Period98}, Sperl 1998) in
single-star mode.

\item {\bf I.} The best fit so far is subtracted from the data. At the
first instance a constant is subtracted. The residual is inputted to
{\it fasper}. If the significance of the peak is above the cut-off, or
already $n_{\rm max}$ periods have been determined, goto {\bf II.}

In single-star mode the frequency sometimes has to be set to twice or
half the frequency found by {\it fasper} at this stage in order to let
the program converge to the correct result.

\item First guesses for the amplitudes (see Eq.~1) are determined, and
then the data plus parameters are inputed to {\it mrqmin}. At least
two calls to {\it mrqmin} are made, and these calls are then repeated until
subsequent $\chi^2$ values agree to within 1\% (or until a maximum of
10 iterations is reached). The $\chi^2$ value is stored. 

\item For the improved fit, for each data point the deviation in terms
of sigma is determined and if the data point with the largest
deviation deviates more than 10$\sigma$ from the fit, the error bar of
that point is set to a very large value, so that that point is ignored
in the further analysis.

\item goto {\bf I.}

\item {\bf II.} Plot the light curves including the best fit, and
write all relevant output to file.

\item Do the final correlation with \DE\ and \M\ using a search radius
of 1.5\arcsec. This takes into account the typical accuracy of the
coordinates in the \OG, \DE\ and \M\ surveys. Multiple matches are
allowed for. The relevant data is written to file.

\end{itemize}

\section{Step~2: Selecting AGB stars and general statistics}

In the second code a preliminary list of LPV candidates is determined,
by applying (if one wishes) selections in magnitude, periods and
amplitudes and by eliminating known non-LPVs and correlating with
known LPVs and/or AGB stars. These issues are discussed here.

\subsection{Known sources}

The advantage of compiling a list of variable objects in the direction
of the MCs that are known not to be LPVs is twofold. First, it
immediately limits the number of sources to be inspected visually in
Step~3. Even more importantly, knowing the light curves of known
non-LPVs helps identifying other such kind of objects in Step~3.

The literature was scanned for lists of objects identified in \OG\ and
other microlensing surveys {\sc macho}, {\sc moa}, {\sc eros} in the
direction of the MCs (hence not restricted to the \OG\ fields).
Eclipsing binaries (EBs) were included from Udalski et al. (1998; OGLE, 1459
in SMC), Wyrzykowski et al. (2003; OGLE, 2580 in LMC), Alcock et
al. (1997; MACHO, 637 in LMC), Bayne et al. (2002; MOA, 167 in SMC),
Grison et al. (1995; EROS, 79 in LMC).
RV Tau objects from Alcock et al. (1998; MACHO, 33 in LMC).
R CrB stars from Alcock et al. (2001; MACHO, 17 in LMC).
A few known and many new candidate QSO from Eyer (2002; OGLE, 133 in SMC
plus LMC), and Geha et al. (2003; MACHO, 59 SMC + LMC).
Blue variable objects (possibly related to the Be phenomenon) 
from Mennickent et al. (2002; OGLE, 1056 in SMC), 
Mennickent et al. (2003; OGLE, 30 in SMC plus LMC), 
Eyer (2002; OGLE, 36 in SMC plus LMC), 
Keller et al. (2002; MACHO, 1280 in LMC).
RR Lyrae stars from Soszy\'nski et al. (2002, OGLE, 556 in SMC; 2003,
7612 in LMC) and Alcock et al. (2000; MACHO, 283 in LMC).
Finally, Cepheids from Afonso et al. (2003; EROS, 880 in SMC and
LMC), second overtone (SO) cepheids from Alcock et al. (1999; MACHO,
47 in LMC) and fundamental mode (FU), first overtone (FO), SO and
double-mode (FU/FO and FO/SO) cepheids from Udalksi et al. 
(1999a,b,c,d; OGLE, 3492 in SMC + LMC).

This amounts to a total of 20863 objects (including double entries
coming from different sources).

In Step~3 the {\sc simbad} database is queried and so previous
observations will normally be identified in this way. On the other
hand, not all (recent) data is yet included there. Therefore a list of
known or suspected LPVs, mostly from recent large survey work, was
compiled.

Two of the largest `old' surveys for LPVs are those by Hughes (1989)
and Reid, Hughes and Glass (1995). I was not able to trace the
relevant tables in these papers in computer readable form, but through
P.~Wood I obtained a table originally prepared by S.~Hughes that seems
to combine the data from these two papers. Eliminating double entries
it contains information on 1317 LPVs in the LMC (I will refer to this
list as the Hughes-list from now on).

From the {\sc moa} survey the 146 LPVs in the LMC discussed by Noda et
al. (2002) are considered. 

M.~Schultheis kindly made available the unpublished periods,
amplitudes, magnitudes of the about 470 LPVs in the LMC bar discovered
by the {\sc agaperos} survey and discussed in detail by Lebzelter et al. (2002)

From Cioni et al. (2003) the 458 objects in the SMC are considered
that have been detected in a mini-survey with {\sc isocam} (Loup et
al. 2004, in prep.), have reliable \DE\ and \M\ counterparts (in the
2nd incremental data release) and have a {\sc macho} lightcurve.

P.~Wood kindly made available the {\sc macho} id numbers, coordinates,
periods and photometry of the 1560 objects he studied in Wood et al. (1999) 
and Wood (2000) in the bar of the LMC.

From Feast et al. (1989), using additional information provided in
Glass \& Lloyd-Evans (2003) and references therein, 58 LMC objects are
considered that {\it de facto} have been defined the Mira $PL$-relation.

Finally, a list of 45 IRAS sources was compiled from Wood et al. (1992),
Wood (1998), Whitelock et al. (2003), that have been monitored in the
infrared and were found to have well determined periods.

The total list of known LPVs contains 4053 objects. 


In addition a list of spectroscopically confirmed M-, S-, and C-stars
and supergiants with accurate coordinates (listed to 1\arcsec\ or
better) was compiled from lists in Westerlund et al. (1981; WOH),
Prevot et al. (1983; PMMR), Wood et al. (1985; WBP), Sanduleak (1989;
SkKM), Reiberot et al. (1993; RAW), Kontizas et al. (2001; KDM),
Morgan \& Hatzidimitriou (1995; MH), Demers et al. (1993), Kunkel et
al. (1997), Groenewegen \& Blommaert (1998; GB98), Kunkel et
al. (2000), Loup et al. (2003, specifically dealing with the
Blanco fields as discussed in Blanco et al. (1980; BMB) and Frogel \&
Blanco 1990) and Cioni et al. (2001), for a total of 13175 stars
(including some double-entries).

As for some of the stars in the Hughes-list a spectral type has been
determined, in fact 13451 stars (including double entries) in this list
have a spectral type assigned. Correlating this list with itself,
using a 4\arcsec\ search radius, reveals 12631 unique entries (2899 C,
19 M, 0 S in the SMC, 8117 C, 1580 M, 16 S in the LMC).  This list is
strongly biased towards carbon stars because of the very large surveys
in both LMC and SMC (notably RAW and KDM), and the lack of similar
surveys for M-stars.

\subsection{Cuts in magnitude, amplitude and period}

In the present paper no cuts on magnitude, amplitude or period have
been made, as the sample discussed below will be restricted to
spectroscopically confirmed M-, S-, C-stars. For reference, in
Groenewegen (2004) the following cuts were applied to the SMC data to
select the LPV candidates: mean OGLE $I < 16.8$ mag, any of the fitted
periods $>50$ days, and any of the fitted amplitudes (in the classical
sense, $C_i = \sqrt {A_i^2 + B_i^2}$ from Eq.~1) $>0.05$ mag.

In hindsight, it would have been preferable to have had a faint limit
(maybe at $I \sim$ 18-18.5) imposed as quite a few spectroscopically
confirmed M,S,C-stars were positionally matched with very faint (mean
$I$ \more 19) \OG\ objects which were often barely variable, and
clearly not the counterpart of the late-type stars.

\subsection{More details}

The content of the code is now described in detail.

\begin{itemize}

\item Read in the output file of Step~1, and the files with the
coordinates of the known non-LPVs, and the known LPVs and
spectroscopically confirmed M-,S-,C-stars.

\item Only periods with an accuracy better than 2\% are retained.

\item The selection on mean $I$-magnitude, amplitude and period may be
carried out. In the present paper no cuts have been applied.

\item Objects that appear twice in different \OG\ fields are
identified and the one with the lowest $\chi^2$ as determined in Step~1
is retained.

\item The remaining objects are correlated on position (search radius
2\arcsec) with the known non-LPVs. The identifiers of both the known
non-LPVs and the remaining objects are written to file.  This
separation assumes that the classification in the literature of an
object as Cepheid, RR Lyrae, eclipsing binary, etc, is correct.

\item The remaining objects are correlated on position (search radius
4\arcsec) with the known LPVs and spectroscopically confirmed
M-,S-,C-stars. The identifiers of the cross-correlated objects are written
to file in Latex format.

In fact, to allow for small differences in the astrometry, corrections
have been determined and applied, as discussed later.

\item Independent of the selection of potential LPVs, in the second
part of the code, general properties are investigated, like the
positional match between \OG\ and \M, and \OG\ and \DE\ sources, and the
difference between $I_{\rm ogle}$ and $I_{\rm denis}$, $J_{\rm 2mass}$
and $J_{\rm denis}$, and $K_{\rm 2mass}$ and $K_{\rm denis}$.

\end{itemize}

\section{Step~3: Visual inspection and literature study}

This last step is time consuming and involves several checks.

\begin{itemize}

\item A visual inspection of the fit to the data. If needed the fit is
re-done with the code in single-star mode, and usually involves
``twisting'' the first frequency it finds to half or twice its
value. On some occasions the data is analysed with an independent
code, {\sc Period98} (Sperl 1998).

\item Cross-correlation with the {\sc simbad} database. The coordinate
list with candidate LPVs is send to the batch queue server of {\sc simbad}.

\item Checking and completing of automatically generated tabular
material in Latex format.

\end{itemize}


\begin{thebibliography}{}

\bibitem[]{} Alard C., Blommaert J.A.D.L., Cesarsky C., et al., 2001, ApJ 552, 289

\bibitem[]{} Alcock C., Allsman R.A., Alves D.R., et al., 1997, AJ 114, 326  

\bibitem[]{} Alcock C., Allsman R.A., Alves D.R., et al., 1998, AJ 115, 1921  

\bibitem[]{} Alcock C., Allsman R.A., Alves D.R., et al., 1999, ApJ 511, 185  

\bibitem[]{} Alcock C., Allsman R.A., Alves D.R., et al., 2000, ApJ 542, 257  

\bibitem[]{} Alcock C., Allsman R.A., Alves D.R., et al., 2001, ApJ 554, 298  

\bibitem[]{} Afonso C., Albert J.N., Alard C., et al., 2003, preprint 

\bibitem[]{} Bayne G., Tobin W., Pritchard J.D., et al., 2002, MNRAS, 331, 609 

\bibitem[]{} Bedding T.R., Zijlstra A.A., Jones A., Foster G., 1998, MNRAS 301, 1073

\bibitem[]{} Bessell M.S., Wood P.R., Lloyd Evans T., 1983, MNRAS 202, 59 (BWL)

\bibitem[]{} Blanco V.M., McCarthy M.F., Blanco B.M., 1980, ApJ 242, 938 (BMB)

\bibitem[]{} Blanco V.M., McCarthy M.F., 1990, AJ 100, 674 (BM)

\bibitem[]{} Blanco V.M., Richer H.B., 1979, PASP 91, 659

\bibitem[]{} Castilho B.V., Gregorio-Hetem J., Spite F., Spite M., Barbuy B., 1998, A\&AS 127, 139

\bibitem[]{} Cioni M.-R.L., Blommaert J.A.D.L., Groenewegen M.A.T., et al.,  2003, A\&A 406, 51 (CBG03)

\bibitem[]{} Cioni M.-R.L., Loup C., Habing H.J., et al., 2000, A\&AS 144, 235 (DCMC)

\bibitem[]{} Cioni M.-R.L., Marquette J.-B., Loup C., et al., 2001, A\&A 377, 945 (CML)

\bibitem[]{} Cutri R.M., Skrutskie M.F., Van Dyk S., 
et al., 2003, Explanatory Supplement to the 2MASS All-Sky Data Release

\bibitem[]{} Delmotte N., Loup C., Egret D., Cioni M.-R., Pierfederici F, 2002, A\&A 396, 143

\bibitem[]{} Demers S., Irwin M.J., Kunkel W.E., 1993, MNRAS 260, 103 (DIK)

\bibitem[]{} Draine B.T., 2003, ARA\&A 41, 241

\bibitem[]{} Egan M.P., Van Dyk S.D., Price S.D., 2001, AJ 122, 1844 (EDP) 

\bibitem[]{} Epchtein N., Deul E., Derriere S., 
et al., 1999, A\&A 349, 236

\bibitem[]{} Eyer L, 2002, AcA 52, 241 

\bibitem[]{} Feast M.W., Glass I.S., Whitelock P.A., Catchpole R.M., 1989, MNRAS 241, 375 (FGWC)

\bibitem[]{} Feast M.W., Whitelock P.A., 1992, MNRAS 259, 6 (FW92) 

\bibitem[]{} Fluks M.A., Plez B., Th\'e P.S., et al., 1994, A\&AS 105, 311 

\bibitem[]{} Frogel J.A., Blanco V.M., 1990, ApJ 365, 168 (FB90)

\bibitem[]{} Frogel J.A., Mould J., Blanco V.M., 1990, ApJ 352, 96 (FMB)

\bibitem[]{} Geha M., Alcock C., Allsman R.A., et al., 2003, AJ 125, 1  

\bibitem[]{} Glass I.S., 1979, MNRAS 186, 317 

\bibitem[]{} Glass I.S., 1986, MNRAS 221, 879

\bibitem[]{} Glass I.S., Lloyd-Evans T., 1981, Nature 291, 303

\bibitem[]{} Glass I.S., Lloyd-Evans T., 2003, MNRAS 343, 67 (GLE)
\
\bibitem[]{} Glass I.S., Schultheis M., 2002, MN 337, 519

\bibitem[]{} Glass I.S., Whitelock P.A., Catchpole R.M., Feast M.W., 1995, MNRAS 273, 383

\bibitem[]{} Grison P., Beaulieu J.-P., Pritchard J. D., et al., 1995,
A\&AS 109, 447 

\bibitem[]{} Groenewegen M.A.T., 2004, in: ``IAU Colloquium 193:
Variable Stars in the Local Group'', eds. D.W. Kurtz \& Karen Pollard,
ASP Conf. Ser., in press

\bibitem[]{} Groenewegen M.A.T., Blommaert J.A.D.L., 1998, A\&A 332, 25 (GB98)

\bibitem[]{} Groenewegen M.A.T., Whitelock P.A., 1996, MNRAS 281, 1347

\bibitem[]{} Hatzidimitriou D., Morgan D.H., Cannon R.D., Croke B.F.W., 2003, MNRAS 341, 1290 (HMCC)


\bibitem[]{} Hughes S.M.G., 1989, AJ 97, 1634

\bibitem[]{} Hughes S.M.G., Wood P.R., 1990, AJ 99, 784 

\bibitem[]{} Ita Y., Tanab\'e T., Matsunaga N., et al., 2004a, MNRAS 347, 720

\bibitem[]{} Ita Y., Tanab\'e T., Matsunaga N., et al., 2004b, MNRAS in press (astro-ph/0312079)

\bibitem[]{} Ita Y., 2004, in: ``IAU Colloquium 193:
Variable Stars in the Local Group'', eds. D.W. Kurtz \& Karen Pollard,
ASP Conf. Ser., in press

\bibitem[]{} Keller S.C., Bessell M.S., Cook K.H., Geha M., 
Syphers D., 2002, AJ 124, 2309 

\bibitem[]{} Kiss L.L., Bedding T., 2003, MNRAS 343, L79

\bibitem[]{} Kiss L.L., Bedding T., 2004, MNRAS 347, L83

\bibitem[]{} Kontizas E., Dapergolas A., Morgen D.H., Kontizas M.,
2001, A\&A 369, 932 (KDM)

\bibitem[]{} Kunkel W.E., Irwin M.J., Demers S., 1997, A\&AS 122, 463 (KID97)

\bibitem[]{} Kunkel W.E., Demers S., Irwin M.J., 2000, AJ 119, 2789 (KDI00)

\bibitem[]{} Lebzelter T., Schulteis M., Melchior A.L., 2002, A\&A 393, 573

\bibitem[]{} Loidl R., Lan\c{c}on A., Jorgensen, U. G., 2001, A\&A 371, 1065

\bibitem[]{} Loup C., Cioni M.-R., Duc P.-A., Fouqu\'e P., Groenewegen
M.A.T., Azzopardi E., Habing H.J., 2003, A\&A in preparation

\bibitem[]{} Melchior A.-L., Hughes S.M.G, Guibert J., 2000, A\&AS 145, 11

\bibitem[]{} Mennickent R.E., Pietrzy\'nski G., Gieren W., Szwewzyk O.,
2002, A\&A 393, 887 

\bibitem[]{} Mennickent R.E., Pietrz\'ynski G., Diaz M., Gieren W.,
2003, A\&A 399, L47 

\bibitem[]{} Meyssonnier N., Azzopardi M., 1993, A\&AS 102, 451 (MA93)

\bibitem[]{} Morgan D.H., Hatzidimitriou D., 1995, A\&AS 113, 539 (MH) 

\bibitem[]{} Munari U, Zwitter T., 2002, A\&A 383, 188 

\bibitem[]{} Nishida S., Tananbe T., Nakada Y., et al.,
2000, MNRAS 313, 136

\bibitem[]{} Noda S., Takeuti M., Abe F., et al., 2002, MNRAS 330, 137

\bibitem[]{} Olivier E.A., Wood P.R., 2003, ApJ 584, 1035

\bibitem[]{} Prevot L., Martin N., Maurice E., Rebeirot E., Rousseeau J., 
1983, A\&AS 53, 255 (PMMR)

\bibitem[]{} Rebeirot E., Azzopardi M., Westerlund B.E., 1993, A\&AS 97, 603 (RAW)

\bibitem[]{} Reid N., Glass I.S., Cathpole R.M., 1988, MNRAS 232, 33 

\bibitem[]{} Reid N., Hughes S.M.G., Glass I.S., 1995, MNRAS 275, 331 

\bibitem[]{} Reid N., Mould J., 1990, ApJ 360, 490 (RM90)

\bibitem[]{} Richer H.B., 1981, ApJ 243, 744

\bibitem[]{} Sanduleak N., 1989, AJ 98, 825 (SkKM)

\bibitem[]{} Sebo K.M., Wood P.R., 1995, ApJ 449, 164

\bibitem[]{} Smith V.V., Plez B., Lambert D.L., Lubowich D.A., 1995, ApJ 441, 735 (SPLL)

\bibitem[]{} Soszy\'nski I., Udalski A., Szyma\'nski M., et al., 2002, AcA 52, 369 

\bibitem[]{} Soszy\'nski I., Udalski A., Szyma\'nski M., et al., 2003, AcA 53, 93 

\bibitem[]{} Sperl M., 1998, Comm. Astroseismology 111, 1

\bibitem[]{} Lloyd Evans T., 1978a, MNRAS 183, 305 (TLE78a)

\bibitem[]{} Lloyd Evans T., 1978b, MNRAS 183, 319 (TLE78b)

\bibitem[]{} Lloyd Evans T., 1980, MNRAS 193, 87 (TLE80)

\bibitem[]{} Trams N.R., Van Loon J.Th., Waters L.B.F.M., 
et al., 1999, A\&A 346, 843 (T99)

\bibitem[]{} Udalski A., Kubiak M., Szyma\'nski M., 1997, AcA 47, 319 

\bibitem[]{} Udalski A., Soszy\'nski I., Szyma\'nski M., 
et al., 1998, AcA 45, 563 

\bibitem[]{} Udalski A., Soszy\'nski I., Szyma\'nski M., 
et al., 1999a, AcA 49, 1 

\bibitem[]{} Udalski A., Soszy\'nski I., Szyma\'nski M., 
et al., 1999b, AcA 49, 45 

\bibitem[]{} Udalski A., Soszy\'nski I., Szyma\'nski M., 
et al., 1999c, AcA 49, 223 

\bibitem[]{} Udalski A., Soszy\'nski I., Szyma\'nski M., 
et al., 1999d, AcA 49, 437 

\bibitem[]{} van Loon J.Th., Zijlstra A.A., Kaper L., et al., 2001, A\&A 368, 239

\bibitem[]{} Wagenhuber J., Groenewegen M.A.T., 1998, A\&A 340, 183 

\bibitem[]{} Westerlund B.E., Olander N., Hedin B., 1981, A\&AS 43, 267 (WOH)

\bibitem[]{} Westerlund B.E., Azzopardi M., Breysacher J., Rebeirot E., 1991, A\&AS 91, 425 (WABR91)

\bibitem[]{} Westerlund B.E., Azzopardi M., Breysacher J., Rebeirot E., 1995, A\&A 303, 107 (WABR95)

\bibitem[]{} Whitelock P.A., Feast M.W., Menzies J.W., Catchpole R.M., 1989, MNRAS 238, 769

\bibitem[]{} Whitelock P.A., Feast M.W., van Loon J.Th., Zijlstra A.A., 2003, MNRAS 342, 86 (WFLZ)

\bibitem[]{} Wilke K., Stickel M., Haas M., et al., 2003, A\&A 401, 873

\bibitem[]{} Wood P.R., 1990, in: ``From Miras to Planetary Nebulae'',
eds. M.O. Mennessier, A. Omont, Editions Frontieres, Gif-sur-Yvette, p. 67

\bibitem[]{} Wood P.R., 1998, A\&A 338, 592

\bibitem[]{} Wood P.R., 2000, PASA 17, 18

\bibitem[]{} Wood P.R., 2003, in: ``Mass-lising pulsating stars and
their circumsetallar matter'', eds. Y. Nakada, M. Honma \& M. Seki,
Kluwer academic publishers, p. 3 

\bibitem[]{} Wood P.R., 2004, in: ``IAU Colloquium 193:
Variable Stars in the Local Group'', eds. D.W. Kurtz \& Karen Pollard,
ASP Conf. Ser., in press

\bibitem[]{} Wood P.R., Alcock C., Allsman R.A., et al., 1999, in:
``IAU Symposium 191: AGB stars'', eds. T. Le Bertre, A. L\`ebre and
C. Waelkens, Kluwer Academic Publishers, ASP, p. 151

\bibitem[]{} Wood P.R., Bessell M.S., Fox M.W., 1983, ApJ 272, 99 (WBF)

\bibitem[]{} Wood P.R., Bessell M.S., Paltoglou G., 1985, ApJ 290, 477 (WBP)

\bibitem[]{} Wood P.R., Sebo K.M., 1996, MNRAS 282, 958

\bibitem[]{} Wood P.R., Whiteoak J.B., Hughes S.M.G., et al., 1992, ApJ 397, 552 

\bibitem[]{} Wood P.R., Zarro D.M., 1981, ApJ 247, 247

\bibitem[]{} Wyrzykowski L., Udalski A., Kubiak M., et al., 2003, AcA 53, 1 

\bibitem[]{} Zebru\'n K., Soszy\'nski I., Wo\'zniak P., et al., 2001, AcA 51, 317

\bibitem[]{} Zijlstra A.A., Bedding T.R., 2003, JAVSO 31, 2

\bibitem[]{} Zijlstra A.A., Bedding T.R., Mattei J.A., 2002, MNRAS 334, 498

\end{thebibliography}
\end{document}